\documentclass[journal, twocolumn]{IEEEtran}

\usepackage{cite}
\usepackage{graphicx}
\usepackage{amsmath}
\usepackage{caption}
\usepackage{booktabs}
\usepackage{subfigure}

\usepackage[norelsize, linesnumbered, ruled, lined, boxed, commentsnumbered]{algorithm2e}      
\SetKwRepeat{Do}{do}{while}%
\SetKw{Kwin}{ in } 
\SetKw{KwTo}{ to }
\SetKw{KwGoto}{goto }

\usepackage{setspace}

\newcommand{\TableRef}[1]{Tab. \ref{#1}}
\newcommand{\FigRef}[1]{Fig. \ref{#1}}

\usepackage{amssymb}   
\usepackage{booktabs}
\usepackage{graphicx}
\usepackage{booktabs}
\usepackage[switch]{lineno}
\usepackage{array}
\usepackage{adjustbox}
\SetAlgoSkip{medskip} 
\usepackage{amsmath}
\usepackage{multirow}
\usepackage{tabularx}
\usepackage{tcolorbox}
\tcbuselibrary{breakable}
\usepackage[hidelinks]{hyperref}

\renewcommand{\color}[1]{}
\begin{document}


\title{Joint Optimization of Trajectory Control, Resource Allocation, and Task Offloading for Multi-UAV-Assisted IoV}

\author{
    Maoxin Ji, Qiong Wu,~\IEEEmembership{Senior Member,~IEEE}, Pingyi Fan,~\IEEEmembership{Senior Member,~IEEE}, Cui Zhang,\\ Nan Cheng, ~\IEEEmembership{Senior Member,~IEEE}, Wen Chen, ~\IEEEmembership{Senior Member,~IEEE}, and Khaled B. Letaief, ~\IEEEmembership{Fellow,~IEEE}

    \thanks{
    	This work was supported in part by Jiangxi Province Science and Technology Development Programme under Grant No. 20242BCC32016, in part by the National Natural Science Foundation of China under Grant No. 61701197, 62531015, and U25A20399, in part by the Basic Research Program of Jiangsu under Grant No. BK20252084, in part by the National Key Research and Development Program of China under Grant No. 2021YFA1000500(4), in part by the Shanghai Kewei under Grant No. 24DP1500500, in part by the Hong Kong Research Grant Council under the Areas of Excellence (AoE) Scheme Grant No. AoE/E-601/22-R and in part by the 111 Project under Grant No. B23008. (Corresponding author: Qiong Wu.)
    	
        Maoxin Ji and Qiong Wu are with the School of Internet of Things Engineering, Jiangnan University, Wuxi 214122, China, and also with the School of Information Engineering, Jiangxi Provincial Key Laboratory of Advanced Signal Processing and Intelligent Communications, Nanchang University, Nanchang 330031, China (e-mail: maoxinji@stu.jiangnan.edu.cn, qiongwu@jiangnan.edu.cn).
        
        Pingyi Fan is with the Department of Electronic Engineering, State Key laboratory of Space Network and Communications, Beijing National Research Center for Information Science and Technology, Tsinghua University, Beijing 100084, China (e-mail: fpy@tsinghua.edu.cn).
        
        Cui Zhang is with the School of Internet of Things Engineering, Wuxi Institute of Technology, Wuxi, 214121, China (e-mail: zhangcui@wxit.edu.cn).
        
        Nan Cheng is with the State Key Lab. of ISN and School of Telecommunications Engineering, Xidian University, Xi'an 710071, China (e-mail: dr.nan.cheng@ieee.org).
        
        Wen Chen is with the Department of Electronic Engineering, Shanghai Jiao Tong University, Shanghai 200240, China (e-mail: wenchen@sjtu.edu.cn).
        
        Khaled B. Letaief is with the Department of Electrical and Computer Engineering, the Hong Kong University of Science and Technology, Hong Kong (email: eekhaled@ust.hk).
        }
}
\maketitle
\begin{abstract}
\textcolor{red}{This paper investigates a multi-Unmanned Aerial Vehicle (UAV) joint base station-assisted Internet of Vehicles (IoV) task offloading system in dense urban environments. To minimize system delay and energy consumption under strict coupling constraints, the complex non-convex optimization problem is decoupled into a hierarchical execution framework. First, a sequential distributed optimization algorithm based on Second-Order Cone Programming (SOCP) is proposed to optimize the 3D flight trajectory of each UAV, ensuring adaptive network coverage. Second, a novel hybrid resource scheduling paradigm synergizing Deep Reinforcement Learning (DRL) and Large Language Models (LLMs) is developed. Within this framework, the DRL agent dictates the initial resource allocation, while the LLM acts as a semantic macro-scheduler to rectify long-tail allocation imbalances for failed and surplus tasks. Crucially, a reward decoupling mechanism is introduced to isolate DRL training from external LLM interventions, thereby ensuring policy convergence. Finally, the task offloading ratios are precisely determined via Linear Programming (LP) within an alternating optimization loop. Simulation results demonstrate that the proposed method significantly outperforms traditional multi-agent reinforcement learning baselines in terms of task success rate and system efficiency.}
\end{abstract}

\IEEEpeerreviewmaketitle

\section{Introduction}\label{intro}

\IEEEPARstart{E}{stablishing} intelligent transportation systems as a critical infrastructure of smart cities heavily relies on advances in Internet of Vehicles (IoV) technologies \cite{17, chu2026v2x, ref001, ref002, ref003, ref004, ref005, ref006}. In IoV scenarios, vehicles acquire information services from roadside units (RSUs) or base stations (BSs) through vehicle-to-infrastructure (V2I) communication, and share real-time traffic data via vehicle-to-vehicle (V2V) communication \cite{18, 19, ref007, ref008, ref009, ref010, ref011, ref012}. However, the explosive growth in data volume and the stringent real-time and reliability requirements of vehicular tasks impose severe challenges on the vehicles' limited onboard computing capabilities \cite{wu2026large}, \cite{xie2025resource}. Mobile Edge Computing (MEC) technology, by offloading computational tasks to edge nodes with more powerful processing capabilities such as RSUs or BSs, emerges as an effective solution to alleviate the computational burden on vehicles \cite{20}.
	
Nonetheless, MEC relying on ground fixed infrastructure faces significant limitations in coping with high-density and highly dynamic IoV environments \cite{xu2026velocity, xu2026enhanced, ref013, ref014, ref015, ref016, ref017, ref018, ref019, ref020}. On one hand, complex urban layouts with dense buildings frequently block communication links between base stations and vehicles, causing severe signal attenuation \cite{777, ref111}. On the other hand, during peak traffic periods, limited computational and wireless communication resources at ground base stations struggle to satisfy the surge of concurrent offloading requests from numerous vehicles, resulting in increased task processing delay and degraded service quality \cite{21}. Moreover, the high mobility of vehicles and the stochastic nature of task arrivals hinder static or semi-static resource allocation and offloading strategies from adapting to the rapidly changing environment.

Recently, rapid development in Unmanned Aerial Vehicle (UAV) technology, especially improvements in payload capacity and endurance, has opened up new possibilities for MEC offloading in IoV. UAVs offer flexible deployment and high maneuverability, enabling them to fly in low-altitude airspace and establish line-of-sight (LoS) communication links with vehicles, effectively overcoming ground obstacles to provide highly reliable communication \cite{9220776, 10233369}. More importantly, UAVs can dynamically adjust their flight trajectories according to real-time vehicle distribution and task load conditions, actively approaching vehicle-dense or high-load areas to provide MEC services. This capability effectively offloads ground base station burden and enhances overall system service capacity.

In the literature, many works have addressed UAV-assisted task offloading and trajectory planning problems in IoV. Existing studies typically focus on specific scenarios. Yan et al.~\cite{10314017} employed deep reinforcement learning (DRL) to optimize task offloading assisted by a single UAV in areas without base stations, aiming to minimize delay via joint trajectory and offloading design. However, assuming fixed-altitude full coverage limits the applicability of their approach to complex IoV scenarios. Liu et al.~\cite{9678115} maximized throughput in resource-constrained scenarios by optimizing UAV trajectories and power allocation under Time Division Multiple Access (TDMA), but the restriction of serving only one vehicle per time slot is too restrictive. Wu et al.~\cite{9541336} proposed a traffic-aware trajectory optimization algorithm using DRL to minimize UAV energy consumption, overlooking communication details. Wang et al.~\cite{10443517} developed a multi-UAV scheduling algorithm to maximize request coverage, simplifying communication and focusing on horizontal trajectories. Chen et al.~\cite{10798447} optimized a multi-UAV MEC system with convex optimization and federated DRL, emphasizing privacy and resource utilization. Liu et al.~\cite{10459245} considered vehicle mobility with a Multi-Agent Deep Deterministic Policy Gradient-based (MADDPG) offloading and migration scheme, optimizing delay and deployment. Nevertheless, none fully considers the critical impact of UAV three-dimensional (3D) flight trajectories, especially altitude variations, on communication coverage quality. In particular, existing works do not consider variations in communication links caused by relative positional changes between UAVs and vehicles.

Research on UAV-assisted MEC for static or low-mobility Internet of Things (IoT) devices provides an important reference foundation for vehicular networks. Reference~\cite{8, 15} focused on multifunctional UAV designs enabling UAVs to act as both MEC servers and relay nodes to improve connectivity. Reference~\cite{9} innovatively proposed a dual-UAV cooperative framework where one UAV handled computation offloading and the other served as a jammer to enhance system security. Reference~\cite{7, 10, 11} dedicated efforts to multi-UAV collaboration optimization, while reference~\cite{13, 14} concentrated on single-UAV systems, jointly optimizing task allocation, bandwidth allocation, computing resource allocation, and UAV trajectory to improve service quality. However, these studies commonly assumed relatively low mobility or fixed user locations (e.g., Reference~\cite{12} assumed fixed user positions and UAV altitude. Reference~\cite{13} assumed full coverage and neglected altitude impact on LoS links. Reference~\cite{14} adopted random walk models. Reference~\cite{15} constrained fixed altitude), which causes its trajectory optimization algorithm to focus primarily on planning the shortest path through all target users, which is difficult to apply in IoV where altitude varies dynamically, randomness is strong, and coverage requirements continuously change, which requires drones to track vehicle positions in real time and rapidly adjust their trajectories to maintain optimal coverage, which directly affects the proportion of vehicles served and system performance, imposing far more stringent demands on the real-time responsiveness and adaptability of trajectory optimization algorithms than in static IoT scenarios.

On the other hand, many existing studies are based on multi-agent deep reinforcement learning (MADRL). However, the policies obtained through MADRL training heavily depend on specific simulation environments, resulting in limited generalization capability \cite{zhang2025drl, gu2025drl}. More importantly, MADRL methods are often regarded as “black boxes” and lack interpretability. In contrast, traditional mathematical optimization methods based on convex optimization have explicit mathematical models and strong interpretability. Nevertheless, in vehicular networks, the joint optimization of UAV flight trajectories, communication resource scheduling, and task offloading usually involves highly non-convex and tightly coupled problems. Such problems are NP-hard and difficult to efficiently solve to global optimality using classical optimization techniques.

In recent years, large language models (LLMs) pretrained on massive datasets have demonstrated powerful general task-solving capabilities and emerging “physical intuition.” LLMs such as Deepseek-R1, ChatGPT, and Gemini can simulate human-like reasoning through chain-of-thought prompting, producing explainable decision rationales that outperform DRL in interpretability. Emerging studies reveal that LLMs can leverage latent physical laws learned from data to solve specific subproblems in communication networks (e.g., power control~\cite{16}) without additional training, achieving performance comparable with specialized DRL algorithms. This opens a novel avenue and framework for tackling complex joint optimization problems in IoV.

In summary, current research exhibits clear limitations in addressing multi-UAV collaborative task offloading problems under high-density and highly dynamic IoV environments: 1) most existing works focus on specific aspects (e.g., trajectory or resource optimization only), lacking comprehensive joint optimization of UAV 3D trajectories, fine-grained communication resource scheduling (power, spectrum), and flexible task offloading strategies (allowing arbitrary proportions of local, UAV, or BS computing); 2) optimization methods confront bottlenecks in interpretability (DRL) or computational complexity (traditional optimization); 3) many studies rely on restrictive assumptions (e.g., fixed altitude, full coverage) and overlook the dynamic changes in communication links caused by UAV trajectory variations. When a vehicle moves out of the UAV's effective coverage area, it will no longer be able to receive services from the UAV, which significantly impacts the overall system performance. Designing trajectory optimization methods that can accurately and effectively ensure coverage of as many vehicles as possible remains a critical challenge. Furthermore, the characteristic of vehicles simultaneously connecting UAVs and BSs further increases the complexity of communication resource management and task scheduling.

To this end, this paper investigates the joint optimization of multi-UAV and BS collaborative task offloading in high-density vehicular networks, aiming to minimize the system’s total delay and weighted total energy consumption\footnote{\href{https://github.com/qiongwu86/Joint-Optimization-of-Trajectory-Cotrol-RA-and-Task-Offloading-for-Multi-UAV-Assisted-IoV}{The source code and LLM prompt have been released at: https://github.com/qiongwu86/Joint-Optimization-of-Trajectory-Control-RA-and-Task-Offloading-for-Multi-UAV-Assisted-IoV}}. The main contributions are summarized as follows:

\begin{enumerate}
	\item Development of the first comprehensive 3D joint optimization model for highly dynamic vehicular networks: This model integrates continuous 3D UAV trajectory planning (including dynamic altitude), joint scheduling of power and spectrum resources, and flexible task offloading among vehicles, UAVs, and base stations. It rigorously incorporates kinematic, communication, processing, and queue stability constraints, addressing the modeling gaps in previous studies.
	\item We propose a convex optimization-based distributed algorithm for fast UAV 3D trajectory planning. By designing load-aware sub-objective functions (combining linear relaxation and penalty functions) and cleverly transforming non-convex motion feasible regions into convex sets, we decouple the complex multi-UAV joint trajectory optimization problem into efficiently solvable subproblems, enabling effective tracking and coverage optimization of vehicle dynamics.
	\item \textcolor{red}{We propose a novel hybrid resource scheduling paradigm that synergizes DRL and LLMs. Within an alternating optimization framework, a DRL agent first dictates the initial joint allocation of resource blocks and transmit power. Subsequently, an LLM acts as a semantic macro-scheduler to resolve long-tail allocation imbalances. By systematically reallocating resources for failed and surplus tasks, the LLM effectively harmonizes system-wide completion times. Driven by few-shot prompting and historical feedback, the LLM iteratively refines its reasoning policy. To ensure edge deployment feasibility, Key-Value (KV) caching is leveraged to bypass redundant computations for highly structured prompts, drastically reducing inference latency. Finally, a deterministic constraint-checking module is integrated to eliminate invalid actions, ensuring strict physical compliance and robust system reliability.}
	\item We propose a linear programming (LP)-based method to determine task offloading proportions. Considering the uncertainty of UAV/BS queue delays at decision epochs, we utilize historical average queue delay data for effective estimation. After fixing UAV trajectories and resource scheduling, the original problem reduces to a deterministic LP with respect to offloading proportions, enabling rapid optimal resolution.
\end{enumerate}

The remainder of this paper is organized as follows. Section~\ref{model} introduces the system model and problem formulation in detail. Section~\ref{method} elaborates on the solution algorithms for the three key subproblems: trajectory planning, resource scheduling, and task offloading. Section~\ref{simulation results} describes the simulation settings, benchmark schemes, and comprehensive performance evaluation results. Section~\ref{conclusion} concludes the paper and discusses future research directions.

\begin{figure*}[t]
	\centering
	\includegraphics[trim=0.5cm 0.5cm 0.5cm 0.5cm, clip, width=0.80\textwidth]{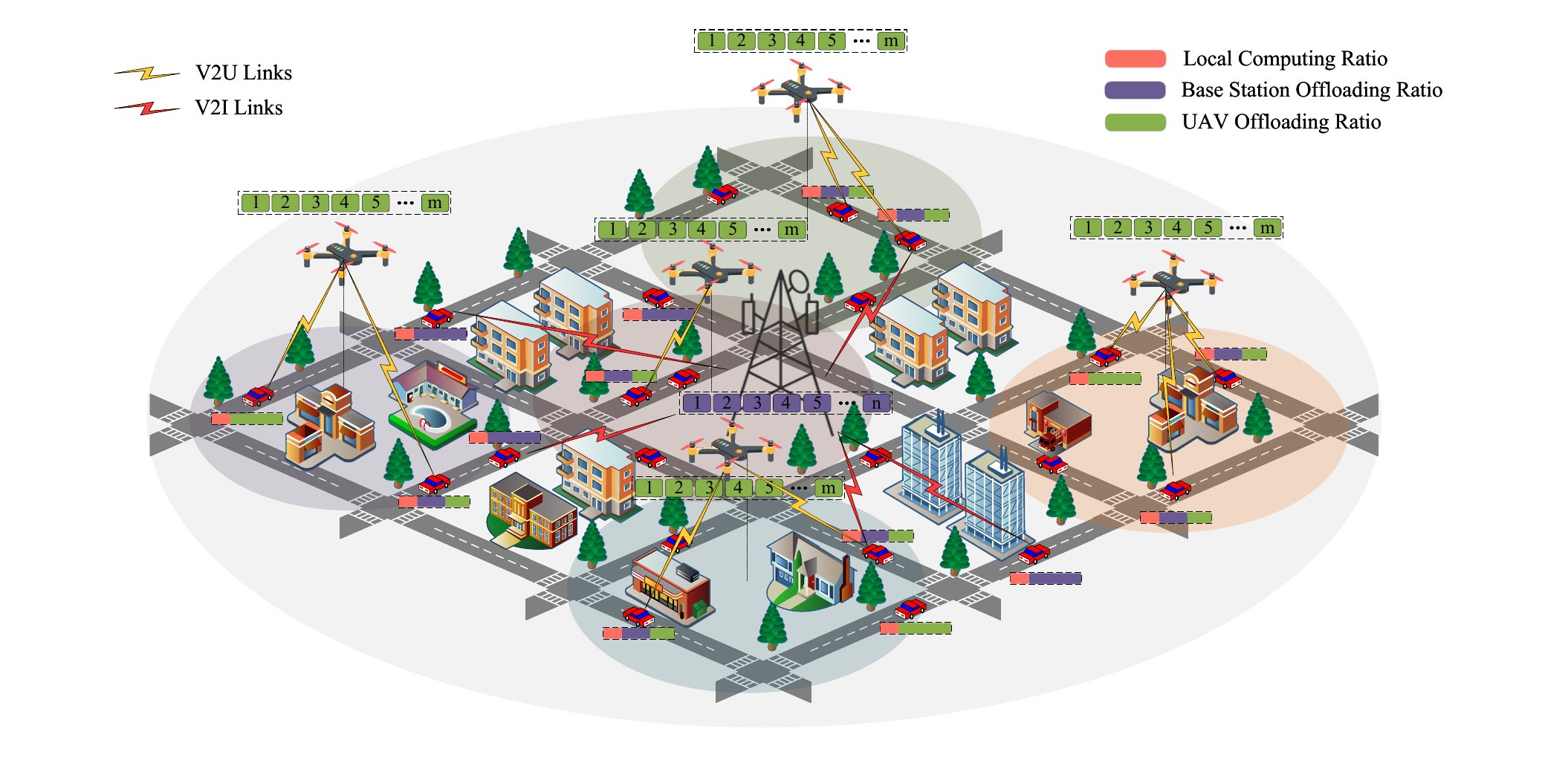}
	\caption{System Model}
	\label{system-model}
\end{figure*}

\section{System Model} \label{model}
As illustrated in Fig.~\ref{system-model}, this paper considers a complex road scenario of size 300m $\times$ 300m, which includes vehicles and UAVs. We denote the set of vehicles in the environment as $\mathcal{M} = \{1, 2, 3, \ldots, M\}$, and the set of UAVs as $\mathcal{U} = \{1, 2, 3, \ldots, U\}$. The UAVs, equipped with edge computing servers, fly at low altitude to provide offloading services to vehicles within their coverage area. A ground BS is located at the center of the environment and serves all vehicles. Both the UAVs and the BS execute computational tasks following a first-in-first-out (FIFO) scheduling policy, each maintaining its own computation queue. Vehicles are initially generated randomly and uniformly distributed across multiple roads within the scenario, moving at a constant speed along the direction of the road. The speeds of different vehicles follow a truncated Gaussian (normal) distribution. When a vehicle reaches an intersection, it randomly selects a feasible direction to turn. To maintain a constant number of vehicles within the simulation, vehicles exiting the boundary re-enter the scenario following traffic regulations.
\vspace{-0.4cm}
\subsection{Mobility Model}
Many existing studies discretize the space into grids to control the flight of UAVs~\cite{23}. Considering the continuity of the three-dimensional space and the realism of the environment, we assume that UAVs can move freely within a limited airspace and spatial range. Let the total simulation time be $T$, which is divided into multiple discrete time slots $\{t_1, t_2, \ldots, t_K\}$. When each time slot is sufficiently small, it can be approximated that the positions of vehicles and UAVs remain fixed within each time slot, while position changes are considered between adjacent time slots~\cite{24}.

At any time slot $t_k$, the three-dimensional position coordinates of an arbitrary vehicle and UAV can be respectively represented as $(x_m(t_k), y_m(t_k), z_m(t_k))$ and $(x_u(t_k), y_u(t_k), z_u(t_k))$. Considering that the UAVs' flight range must be restricted within the simulation scenario, the following constraints apply:
\begin{equation} \label{1}
	\begin{cases}
		x_u^{\min} \leq x_u(t_k) \leq x_u^{\max}, \\
		y_u^{\min} \leq y_u(t_k) \leq y_u^{\max}, \\
		z_u^{\min} \leq z_u(t_k) \leq z_u^{\max},
	\end{cases}
\end{equation}
where $x_u^{\min}$, $x_u^{\max}$, $y_u^{\min}$, $y_u^{\max}$, $z_u^{\min}$, and $z_u^{\max}$ denote the boundaries of the UAVs' allowed flight area.

According to~\cite{6863654}, the flight altitude of a UAV is related to its coverage (effective line-of-sight transmission) range. Suppose the maximum elevation angle of the UAV is $\theta_{\max}$, then the horizontal coverage radius of UAV $u$ at time slot $t_k$ can be expressed as:
\begin{equation}
	R_u(t_k) = z_u(t_k) \tan(\theta_{\max}),
\end{equation}
where $z_u(t_k)$ is the altitude of UAV $u$ at time slot $t_k$. To avoid collisions, multiple UAVs must maintain a certain spatial separation, which should satisfy the following constraint:
\begin{equation} \label{3}
	\| \mathbf{p}_u(t_k) - \mathbf{p}_v(t_k) \| \geq d_{\min}, \quad \forall u \neq v,
\end{equation}
where $\mathbf{p}_u(t_k) = (x_u(t_k), y_u(t_k))$ denotes the horizontal position of UAV $u$ at time slot $t_k$, and $d_{\min}$ is the minimum horizontal distance between UAVs. Due to the limitations of the propulsion system, the changes in horizontal position and altitude of a UAV between adjacent time slots need to satisfy the following constraints:
\begin{equation}  \label{4}
	\begin{cases}
		l_u(t_k) = v_u(t_k) \cdot \cos(\varphi_u(t_k)) \cdot \Delta t \leq l_{\max}^h, \\
		z_u(t_{k+1}) - z_u(t_k) = v_u(t_k) \cdot \sin(\varphi_u(t_k)) \cdot \Delta t \leq l_{\max}^v,
	\end{cases}
\end{equation}
where $l_u(t_k)$ denotes the horizontal displacement of UAV $u$ during time slot $t_k$, $v_u(t_k)$ is the UAV's speed at time slot $t_k$, $\varphi_u(t_k)$ is the elevation angle (flight pitch angle) at time slot $t_k$, $\Delta t$ is the duration of each time slot, and $l_{\max}^h$ and $l_{\max}^v$ are the maximum allowed displacements in the horizontal and vertical directions, respectively. 

\textcolor{red}{To accurately evaluate the operational endurance, this work adopts a power consumption model consistent with rotary-wing aerodynamics \cite{zeng2019energy}. Instead of a simplified linear assumption, the model explicitly describes the non-linear relationship between flight speed and propulsion power. The horizontal propulsion power $P_h$ follows a characteristic U-shaped profile \cite{stolaroff2018energy}, modeled as a function of the horizontal speed $v_{xy}$:}
\begin{equation}
	\textcolor{red}{P_h(v_{xy}) = P_{\mathrm{hover}} \left( \frac{1}{1 + \left(\frac{v_{xy}}{v_{\mathrm{ref}}}\right)^2} \right) + c_d \cdot v_{xy}^3,}
\end{equation}
\textcolor{red}{where $P_{\mathrm{hover}}$ denotes the baseline hovering power, $v_{\mathrm{ref}}$ represents the induced power decay factor, and $c_d$ is the aerodynamic drag coefficient. }

\textcolor{red}{For vertical maneuvering, the asymmetry of gravity work dictates that ascent requires overcoming gravitational potential while descent allows for limited energy recovery. The vertical power $P_v(v_z)$ is defined as:}
\begin{equation}
	\textcolor{red}{P_v(v_z) = 
	\begin{cases} 
		mg v_z + c_v v_z^2, & v_z > 0, \\
		\alpha \cdot mg v_z, & v_z \le 0,
	\end{cases}}
\end{equation}
\textcolor{red}{where $v_z$ is the vertical velocity (positive upwards) and $\alpha \in (0,1)$ is the energy reduction coefficient during descent. Consequently, combining the ancillary power $P_{\mathrm{anc}}$ of onboard electronics, the total energy consumption $E^u(t_k)$ over a time slot $t_k$ is calculated as:}
\begin{equation} \label{77777}
	\textcolor{red}{E^u(t_k) \approx \left( P_h(v_{xy, t_k}) + P_v(v_z, t_k) + P_{\mathrm{anc}} \right) \cdot \Delta t.}
\end{equation}

\vspace{-0.2cm}
\subsection{Communication Model}

This paper considers a complex urban environment, where all vehicles are assumed to be within the coverage area of the BS. The BS establishes communication with vehicles through V2I links. The three-dimensional coordinates of the base station are denoted as \(\mathbf{b} = (x_I, y_I, z_I)\).

According to 5G NR-V2X technology, vehicles communicate with the BS via V2I links, and with UAVs via ground-to-air (G2A) links. It is assumed that the V2I and G2A communication links share the same frequency bandwidth \(B\), which is divided into different resource blocks in both frequency and time slots.

Therefore, each V2I and G2A link transmits data over orthogonal frequency bands corresponding to their allocated bandwidths \(B_m^I(t_k)\) and \(B_m^{u}(t_k)\), respectively, within their corresponding time slots. These satisfy the following constraint:

\begin{equation} \label{7}
	\sum_{m=1}^M B_m^I(t_k) + \sum_{u=1}^U \sum_{m=1}^M \alpha_m^u(t_k) B_m^{u}(t_k) \leq B,
\end{equation}

where
\begin{equation}
\alpha_m^u(t_k) = 
\begin{cases}
	1, & d_m^{h,u}(t_k) \leq R_u(t_k), \\
	0, & d_m^{h,u}(t_k) > R_u(t_k),
\end{cases}
\end{equation}
where \(d_m^{h,u}(t_k)\) denotes the horizontal distance between UAV \(u\) and vehicle \(m\) at time slot \(t_k\), and \(\alpha_m^u(t_k)\) indicates whether vehicle \(m\) is within the coverage range \(R_u(t_k)\) of UAV \(u\) at time slot \(t_k\). Due to the wireless signal of V2I communication experiencing free-space path loss, shadow fading, and fast fading, taking into account the transmit antenna gain of the vehicle and the receive antenna gain of the base station, the received power at the base station from vehicle \(m\) at time slot \(t_k\) can be described as:

\begin{equation}
	\begin{aligned}
		P^{I}_{rx,m}(t_k) &= P^{I}_{tx,m}(t_k) - L^{I}_{m}(t_k) + S_m(t_k) \\
		&+ F^{I}_{m,r}(t_k) + G_m^t + G_I^r, \quad \forall k
	\end{aligned}
\end{equation}
where \(P^{I}_{tx,m}(t_k)\) denotes the transmit power of vehicle \(m\) communicating with the base station at time slot \(t_k\), \(L^{I}_{m}(t_k) = 128.1 + 37.6 \times \log_{10} \frac{d^{I}_{m}(t_k)}{1000}\) represents the path loss. Here, the distance is calculated as \(d^{I}_{m}(t_k) = \sqrt{(x_m(t_k) - x_I)^2 + (y_m(t_k) - y_I)^2 + (z_m - z_I)^2}\), where \(z_m\) denotes the common antenna height of all vehicles.

The shadow fading is modeled as \(S_m^{t_k} = e^{-\frac{\Delta d_m}{D_c}} S_m(t_{k-1}) + \sqrt{1 - e^{-\frac{2 \Delta d_m}{D_c}}} \ \mathcal{N}(0, \sigma_{s})\), where \(\Delta d_m\) represents the moving distance of vehicle \(m\) within a single time slot, \(D_c\) denotes the decorrelation distance, and \(\sigma_{s}\) is the standard deviation of shadow fading.
The fast fading is expressed as \(F^{I}_{m,r}(t_k) = 20 \log_{10} |h_{m,r}(t_k)|\), where \(h_{m,r}(t_k)\) is the Rayleigh distributed random variable representing the fast fading of vehicle \(m\) on the allocated resource block \(r\).
\(G_m^t\) denotes the transmit power gain of vehicle \(m\), and \(G_I^r\) represents the receive power gain of the base station. Based on this, the data rate of V2I communication can be expressed as:

\begin{equation} \label{eq:V2I_rate}
	R_m^I(t_k) = B_m^I(t_k) \log_2 \left(1 + \frac{P_{total,m}^I(t_k)}{N_m^I(t_k)} \right), \quad \forall t_k
\end{equation}
where $P_{total,m}^I(t_k) = 10^{\frac{P_{rx,m}^I(t_k)}{10}}$ denotes the linear-scale transmission power of vehicle \(m\) communicating with the base station at time slot \(t_k\), where \(N_m^I(t_k)\) represents the linear-scale noise power at time slot \(t_k\).

The UAV provides task offloading services to vehicles in the air. Since the UAV operates in an open-sky environment, G2A communication has relatively high LoS connectivity \cite{25}. Based on the path loss model \cite{6863654}, at time slot \(t_k\), the probability that vehicle \(m\) has LoS transmission with UAV \(u\) can be calculated as follows:

\begin{equation} \label{eq:LoS_probability}
	P_{m,u}^{LoS}(t_k) = \frac{1}{1 + \omega_a \exp\left( -\omega_b \left( \arcsin \frac{z_u(t_k)}{d_{m,u}(t_k)} - \omega_a \right) \right)},
\end{equation}
where

\begin{equation} \label{eq:distance_uav_vehicle}
	d_{m,u}(t_k) = \sqrt{ \left( \Delta x(t_k) \right)^2 + \left( \Delta y(t_k) \right)^2 + \left( \Delta z(t_k) \right)^2 },
\end{equation}
where \(\Delta x(t_k) = x_m(t_k) - x_u(t_k)\) denotes the difference in the \(x\)-coordinates between UAV \(u\) and vehicle \(m\). The terms \(\Delta y(t_k)\) and \(\Delta z(t_k)\) are defined similarly for the \(y\) and \(z\) axes, respectively. \(\omega_a\) and \(\omega_b\) are constants describing the propagation environment characteristics. It can be seen that the LoS probability between the vehicle and UAV depends on their relative positions and heights. The LoS probability depends on the elevation angle between the vehicle and the UAV. Correspondingly, the probability of non-line-of-sight (NLoS) propagation can be expressed as:  
\begin{equation} \label{eq:p}
	P_{m,u}^{\text{NLoS}}(t_k) = 1 - P_{m,u}^{\text{LoS}}(t_k).
\end{equation}

The path loss during LoS and NLoS transmission can be expressed as:

\begin{equation} \label{eq:pathloss_LoS}
	C_{m,u}^{\text{LoS}}(t_k) = C_{m,u}^{\text{FS}}(t_k) + \eta_{\text{LoS}},
\end{equation}

\begin{equation} \label{eq:pathloss_NLoS}
	C_{m,u}^{\text{NLoS}}(t_k) = C_{m,u}^{\text{FS}}(t_k) + \eta_{\text{NLoS}},
\end{equation}
where $ C_{m,u}^{\text{FS}}(t_k) = 20 \log_{10} ( \frac{4 \pi d_{m,u}^u(t_k) f_v}{V_c}),$  and $\eta_{LoS}$ denotes the excess path loss associated with LoS conditions, $\eta_{NLoS}$ denotes the excess path loss for NLoS, and $f_v$ represents the carrier frequency for G2A communication.  
Therefore, the average G2A path loss can be represented as:  
\begin{equation} \label{eq:average_pathloss}
	C_{m}^{u}(t_k) = C_{m,u}^{FS}(t_k) + P_{m,u}^{LoS}(t_k) \eta_{LoS} + P_{m,u}^{NLoS}(t_k) \eta_{NLoS}.
\end{equation}
Substituting $P_{m,u}^{NLoS}(t_k) = 1 - P_{m,u}^{LoS}(t_k)$ into the above expression, it can be simplified as:
\begin{equation}
	C_{m}^{u}(t_k) = C_{m,u}^{FS}(t_k) + P_{m,u}^{LoS}(t_k) \big(\eta_{LoS} - \eta_{NLoS}\big) + \eta_{NLoS}.
\end{equation}

Since the vehicle is in high-speed motion, according to the Lais fall model, the channel shadowing of vehicle \(m\) at time slot \(t_k\) and resource block \(r\) can be calculated as:$F_{m,r}(t_k) = 20 \log_{10} \left( \sqrt{\frac{K}{K+1}} e^{j \theta_{m,r}(t_k)} + \sqrt{\frac{1}{K+1}} \sqrt{\frac{1}{2}} \big(N^{(0,1)} + j N^{(0,1)}\big) \right)$. Therefore, the received power can be expressed as:
\begin{equation}
	P_{r,m}^{u}(t_k) = P_{tx,m}^{u}(t_k) - C_{m}^{u}(t_k) - F_{m,r}(t_k) + G_{m}^{t} + G_{u}^{r},
\end{equation}
where \(P_{tx,m}^{u}(t_k)\) denotes the transmit power of vehicle \(m\) to UAV \(u\), and \(G_{u}^{r}\) represents the UAV \(u\)'s receiving antenna gain.
Based on this, the transmission rate between vehicle \(m\) and UAV \(u\) within the coverage area can be expressed as:

\begin{equation}
	R_{m,r}^{u}(t_k) = B_{m}^{u}(t_k) \log_2 \left( 1 + \frac{P_{m}^{u}(t_k)}{N_{m,r}^{u}(t_k)} \right),
\end{equation}
where $P_{m}^{u}(t_k) = 10^{\left(\frac{P_{tx,m}^{u}(t_k)}{10}\right)}$, $B_{m}^{u}(t_k)$ denotes the transmission bandwidth, and \(N_{m,r}^{u}(t_k)\) represents the noise power linear value.
\vspace{-0.2cm}
\subsection{Task Offloading Strategy}
In the considered environment, the total task volume each vehicle needs to offload at time \(t_k\) is given by $D(t_k) = \{D_1(t_k), D_2(t_k), \ldots, D_M(t_k)\}$. All vehicles are equipped with on-board units (OBUs) that have identical computing frequency \(f_m\). All UAVs carry lightweight aerial servers with computing frequency \(f_u\). The base station hosts high-performance servers with computing frequency \(f_I\).

\textcolor{red}{Each vehicle can request service from only one UAV. When covered by multiple UAVs at time $t_k$, instead of relying solely on the computational load, vehicle $m$ intelligently selects the target UAV by evaluating a joint channel-aware and load-aware cost metric. Specifically, the selection criterion is formulated to minimize the weighted cost $\Psi_{m,u}(t_k) = \lambda_1 D_u(t_k) + \lambda_2 C_{m}^{u}(t_k)$, where $D_u(t_k)$ denotes the total task load of all vehicles within the coverage area of UAV $u$, $C_{m}^{u}(t_k)$ is the real-time average G2A path loss derived in Eq. (\ref{eq:average_pathloss}), and $\lambda_1, \lambda_2$ are normalized importance weights. Vehicle $m$ associates with the UAV $u^*$ that yields the minimum $\Psi_{m,u}(t_k)$. Task proportions are independent over time slots, allowing vehicles to offload any portion of their tasks to edge servers.}
Thus, the offloading strategy for vehicle \(m\) at time \(t_k\) can be expressed as:
\begin{equation}\label{eq:task_offloading}
	\textcolor{red}{\gamma_m^o(t_k) + \alpha_m^u(t_k) \gamma_m^u(t_k) + \gamma_m^I(t_k) = 1,}
\end{equation}
where, \(\gamma_m^o(t_k)\) and \(\gamma_m^l(t_k)\) denote the proportions of tasks computed locally and offloaded to the base station, respectively; \(\gamma_m^u(t_k)\) is the proportion offloaded to the UAV. The indicator \(\alpha_m^u(t_k)\) equals 1 if vehicle \(m\) is covered by at least one UAV, and 0 otherwise. 
When \(\alpha_m^u(t_k) = 0\), vehicle \(m\) is outside UAV coverage and cannot offload tasks to UAVs; otherwise, at least one UAV can serve vehicle \(m\).

Based on the above task partitioning, the computation delay of task offloading can be calculated as follows:
\begin{equation}
	\begin{cases}
		T_{m}^{o}(t_k) = \dfrac{\gamma_{m}^{o}(t_k) D_{m}(t_k) c}{f_m}, \\
		T_{m}^{u}(t_k) = \dfrac{\gamma_{m}^{u}(t_k) D_{m}(t_k) c}{f_u}, \\
		T_{m}^{I}(t_k) = \dfrac{\gamma_{m}^{I}(t_k) D_{m}(t_k) c}{f_I}, \\
	\end{cases}
\end{equation}
where \(c\) represents the number of CPU cycles required to process one bit of data. The transmission delay for vehicle offloading task can be calculated as:
\begin{equation}
	\begin{cases}
		T_{m}^{V2I}(t_k) = \dfrac{\gamma_{m}^{I}(t_k) D_{m}(t_k)}{R_{m}^l(t_k)}, \\
		T_{m}^{G2A}(t_k) = \dfrac{\gamma_{m}^{u}(t_k) D_{m}(t_k)}{R_{m}^u(t_k)}.
	\end{cases}
\end{equation}

Since each UAV serves multiple vehicles, task execution is assumed to follow a FIFO discipline. Consequently, the queuing delay for vehicle \(m\)'s task is calculated as:
\begin{equation}\label{21}
	T_{m,u}^{que}(t_k) = \max \{T_{m,u}^{ar}(t_k), T_{u}^{last}(t_k)\} - T_{m,u}^{ar}(t_k),
\end{equation}
where \(T_{m,u}^{ar}(t_k)\) represents the time when vehicle \(m\)'s data arrives at UAV \(u\), and \(T_{u}^{last}(t_k)\) represents the current last task processing time in the UAV queue \(u\).
Similarly, the waiting delay at the base station can be expressed as:
\begin{equation}\label{22}
	T_{m}^{que}(t_k) = \max \left\{ T_{m,I}^{ar}(t_k),\ T_{I}^{last}(t_k) \right\} - T_{m,I}^{ar}(t_k),
\end{equation}
where \(T_{m,I}^{ar}(t_k)\) represents the time when data from vehicle \(m\) arrives at the base station, and \(T_{I}^{last}(t_k)\) is the completion time of the last task in the current queue at the base station.

Due to the limited computing capacity of UAVs and limited queue size, the amount of data offloaded to UAV \(u\) at each time slot cannot exceed its computational capacity, i.e.,
\begin{equation} \label{25}
	0 \leq \sum_{m=1}^M \alpha_m^u(t_k) \gamma_{m}^u(t_k) D_{m}(t_k) \leq \frac{f_u \Delta t}{c},
\end{equation}
where \(\Delta t\) denotes the length of one time slot. Similarly, the base station must also impose an upper limit on the offloaded task amount at each time slot:
\begin{equation} \label{26}
	0 \leq \sum_{m=1}^M \gamma_{m}^o(t_k) D_{m}(t_k) \leq \frac{f_I \Delta t}{c}.
\end{equation}
Note that since the queue constraints are imposed within each individual time slot and tasks not completed within the slot are considered failed, the system does not involve queue dynamics across multiple slots. Therefore, there is no need to apply Lyapunov optimization methods, which are typically used to ensure long-term queue stability.
\vspace{-0.4cm}
\subsection{Optimization Problem}
The optimization objective is to minimize both energy consumption and system delay. Given the parallel execution of computation and transmission tasks, the total delay is formulated as:
\begin{equation} \label{T}
	T = \sum_{t_k=1}^{K} \sum_{m=1}^{M} T_m(t_k),
\end{equation}
where:
\begin{equation}
	\begin{aligned}
		T_m(t_k) &= \max \biggl\{ 
		T_{m}^o(t_k), \
		\sum_{u=1}^{U} \alpha_{m}^u(t_k) \bigl( T_{m,u}^{G2A}(t_k) + T_{m,u}^{que}(t_k) \\
		&+ T_{m}^u(t_k) \bigr), \
		T_{m}^{V2I}(t_k) + T_{m}^{que}(t_k) + T_{m}^I(t_k)
		\biggr\}.
	\end{aligned}
\end{equation}

Under ideal conditions, all vehicle computation tasks should be completed within the task deadline, thus the following constraint holds:
\begin{equation}
	T_{m}(t_k) \leq T_{m}^{max}(t_k).
\end{equation}
Considering the resource constraints in practical environments, it is difficult for all vehicles to satisfy this constraint. To ensure fairness in resource competition among vehicles and to avoid that vehicles with loose time requirements occupy resources at the expense of vehicles with tight deadlines, we introduce a delay exceeding penalty \(\xi_m\). Then the delay constraint becomes:
\begin{equation} \label{30}
	T_{m}(t_k) \leq T_{m}^{max}(t_k) + \xi_m(t_k).
\end{equation}
Considering the different delay scales of various vehicle tasks, we transform the optimization objective into normalized delay based on task time constraints. Eq.~\eqref{T} can be rewritten as:
\begin{equation}
	T = \sum_{t_k=1}^{K} \sum_{m=1}^{M} \frac{T_m(t_k)}{T_m^{max}(t_k)}.
\end{equation}
Considering the energy consumption during communication between vehicles, UAVs, and BS, as well as the energy consumed by vehicles, UAVs, and base stations for computation, the transmission energy consumed by vehicles for offloading tasks to the base station and UAV can be expressed as:
\begin{equation}
	\begin{cases}
		E_{m}^{V2I}(t_k) = P_{m}^I T_{m}^{V2I}(t_k), \\
		E_{m}^{G2A}(t_k) = P_{m}^u T_{m}^{G2A}(t_k).
	\end{cases}
\end{equation}
\begin{figure}[t]
	\centering
	\includegraphics[trim=0.5cm 0.5cm 0.5cm 0.5cm, clip, width=0.8\columnwidth]{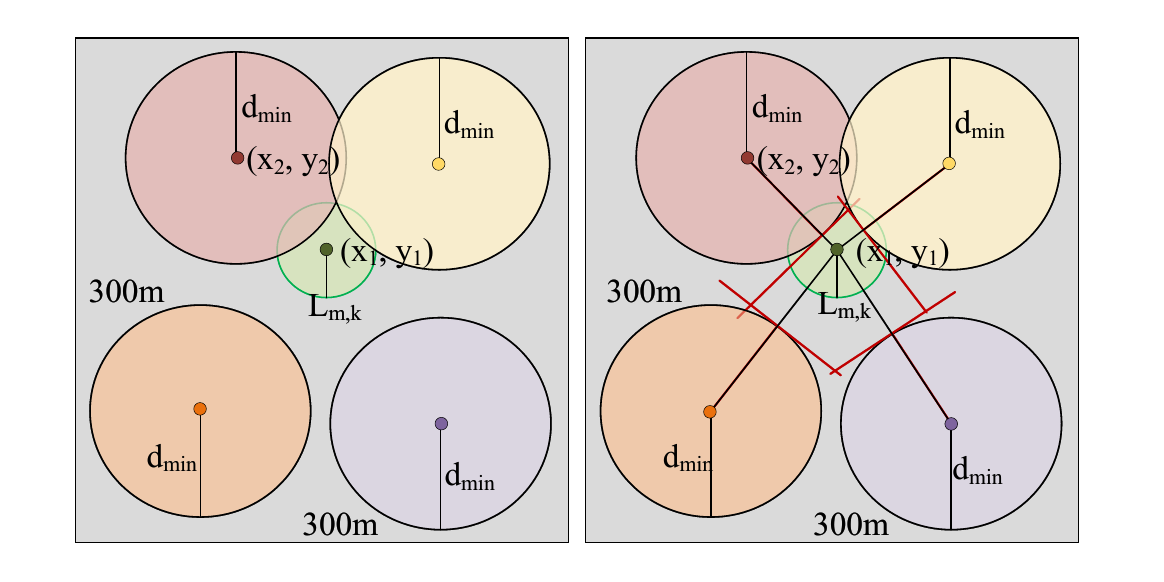}
	\caption{Collision Avoidance Constraint Illustration}
	\label{2}
\end{figure}
The energy consumed by vehicles, UAVs, and base stations for data processing can be calculated by the following formulas:
\begin{equation}
	\begin{cases}
		E_{m}^o(t_k) = \kappa f_{m}^3 T_{m}^o(t_k), \\
		E_{m}^u(t_k) = \kappa f_{u}^3 T_{m}^u(t_k), \\
		E_{m}^I(t_k) = \kappa f_{I}^3 T_{m}^I(t_k),
	\end{cases}
\end{equation}
where \(\kappa \geq 0\) denotes the effective switched capacitance. \textcolor{red}{The flight energy consumption of the UAVs can be calculated according to Eq.~\eqref{77777}.} Therefore, the total system energy consumption can be expressed as:
\begin{equation}
	\begin{aligned}
		&E = \sum_{t_k=1}^{K} \sum_{m=1}^{M} \biggl(
		E_m^o(t_k) + E_m^{V2I}(t_k) + E_m^I(t_k) \\
		&+ \sum_{u=1}^{U} \alpha_m^u(t_k) \bigl( E_m^{G2A}(t_k) + E_m^u(t_k) \bigr) 
		\biggr) + \textcolor{red}{\sum_{t_k=1}^{K}\sum_{u=1}^{U} E^u(t_k).}
	\end{aligned}
\end{equation}
\textcolor{red}{Therefore, by jointly optimizing the UAV's 3D coordinates $\mathcal{C}_u = \bigl(x_u(t_k), y_u(t_k), z_u(t_k)\bigr)$, the vehicle's transmission power $P_m^u(t_k)$ and $P_m^I(t_k)$, the allocation ratios $\gamma_m^{o}(t_k)$, $\gamma_m^{u}(t_k)$, $\gamma_m^{I}(t_k)$ for vehicle tasks, and the resource block allocation $\mathcal{R}$, the objective function can be minimized.} The final optimization problem can be expressed as:
\begin{subequations}\label{eq:optimization}
	\begin{align}
		\min_{\substack{
				\mathcal{C}_u, P \\
				\gamma, \mathcal{R}
		}} 
		&\quad \omega_1 T + \omega_2 E + \omega_3 \sum_{m=1}^{M} \xi_m \label{eq:objective} \\
		\text{s.t.}\quad
		&0 \leq \gamma_m^{o}(t_k) \leq 1, \quad \forall m,  \label{eq:con_a} \\
		&0 \leq \gamma_m^{u}(t_k) \leq 1, \quad \forall m, \forall u,  \label{eq:con_b} \\
		&0 \leq \gamma_m^{I}(t_k) \leq 1, \quad \forall m,  \label{eq:con_c} \\
		&\gamma_m^{o}(t_k) + \alpha_m^{u}(t_k) \gamma_m^{u}(t_k) + \gamma_m^{I}(t_k) = 1,\ \forall m, \forall u,  \label{eq:con_d} \\
		&\textcolor{red}{v_u(t_k) \leq v_{max}},  \label{eq:con_gg}\\
		&0 \leq P_m^{u}(t_k) \leq P_m^{\max}, \quad \forall m, \label{eq:con_e} \\
		&0 \leq P_m^{I}(t_k) \leq P_m^{\max}, \quad \forall m,  \label{eq:con_f} \\
		&\alpha_m^{u}(t_k) \in \{0,1\}, \quad \forall m, \forall u,  \label{eq:con_g} \\
		&\xi_m \geq 0, \quad \forall m \label{penalty} \\
		&\text{Constraints \eqref{1}, \eqref{3}, \eqref{4}, \eqref{7}, \eqref{25}, \eqref{26}, \eqref{30}}.  \label{eq:other_cons}
	\end{align}
\end{subequations}
where \(\omega_1\), \(\omega_2\) and \(\omega_3\) represent the weights of system delay, energy consumption and delay exceeding penalty, respectively. Constraints \eqref{eq:con_a}--\eqref{eq:con_d} limit the offloading ratios of vehicle tasks; constraints \eqref{eq:con_e}--\eqref{eq:con_f} limit the transmission power range of vehicles; constraint \eqref{eq:con_g} indicates whether vehicle \(m\) is covered by UAV \(u\). \textcolor{red}{Eq.~\eqref{1} restricts the UAV's motion space, and constraint \eqref{eq:con_gg} restricts the UAV's max speed.} \textcolor{red}{Eq.\eqref{3} constrains the minimum distance between any two UAVs. Eq.\eqref{4} describes the maximum displacement limit of UAVs between consecutive time slots.} Eq.~\eqref{7} limits the system bandwidth allocation. Eq.~\eqref{25}--\eqref{26} constrain the maximum data amount offloaded to the base station and UAV. Eq.~\eqref{30} restricts the delay upper bound for each vehicle task. Due to strong coupling among variables and integer constraints, this is a highly complex optimization problem. We decompose it into three subproblems to find an approximate optimal solution.
\section{Problem Decomposition and Solution} \label{method}

In this section, we decompose the optimization problem into three subproblems: UAV trajectory planning, resource allocation, and task proportion assignment. By integrating DRL, LLMs, convex optimization, and linear programming methods, approximate optimal solutions are obtained.

\subsection{UAV Trajectory Planning Problem}
\subsubsection{Convexification and Constraint Transformation}
\textcolor{red}{To formulate a computationally tractable convex optimization problem, both the non-convex energy consumption model and the collision avoidance constraints must be appropriately transformed.}

\textcolor{red}{First, the theoretical propulsion power model presented in the system model is fundamentally non-convex. Specifically, the horizontal flight power $P_h(v)$ consists of profile power, induced power, and parasitic power. To incorporate energy awareness into the convex planning framework, we mathematically approximate the non-convex model by applying a second-order Taylor expansion around the hovering state ($v=0$). For the complex induced power term, let $\delta = \frac{v^2}{2v_0^2}$, where $v_0$ is the mean induced velocity in hover. In the low-to-medium speed regime ($\delta \ll 1$), the induced power can be expanded and approximated as:}
\begin{equation}
	\textcolor{red}{\begin{aligned}
		P_{\text{ind}}(v) &= P_i \left( \sqrt{1 + \delta^2} - \delta \right)^{1/2} \approx P_i \left( 1 - \delta \right)^{1/2}\\ 
		&\approx P_i \left( 1 - \frac{1}{2}\delta \right) 
		= P_i - P_i \frac{v^2}{4v_0^2}.
	\end{aligned}}
\end{equation}
\textcolor{red}{This derivation explicitly reveals the translational lift effect, where the induced power decreases quadratically with airspeed. Combining this with the profile power $P_0(1 + \frac{3v^2}{U_{\text{tip}}^2})$ and bounding the cubic parasitic drag $c_d v^3$, the total horizontal power $P_h(v)$ is convexified by constructing a quadratic upper bound:}
\begin{equation} \label{eq:taylor_ph}
	\textcolor{red}{\begin{aligned}
		P_h(v) &\approx \left(P_0 + P_i\right) + \left( \frac{3P_0}{U_{\text{tip}}^2} - \frac{P_i}{4v_0^2} \right) v^2 + c_d v^3 \\
		&\le P_{\text{hover}} + \mu v^2,
	\end{aligned}}
\end{equation}
\textcolor{red}{where $P_{\text{hover}} = P_0 + P_i$ is the baseline hovering power, and $\mu \ge 0$ is a fitted regularization coefficient ensuring the convexity. In a discrete-time framework with time step $\Delta t$, the horizontal energy cost for UAV $i$ is proportional to its squared Euclidean displacement, i.e., $E_{h,i} \propto \| \mathbf{p}_{i} - \mathbf{p}_{i}^{\mathrm{prev}} \|_2^2$, which is strictly convex. }

\textcolor{red}{For vertical maneuvering, the energy consumption exhibits asymmetry: overcoming gravity requires significant power, whereas descending consumes less energy. To capture this physical behavior within a convex formulation, we model the vertical energy cost using a piecewise linear function of the altitude change $\Delta z_i = z_i - z_i^{\mathrm{prev}}$:}
\begin{equation} \label{eq:convex_pv}
\textcolor{red}{	E_{v,i} = w_{\text{up}} [\Delta z_i]^+ + w_{\text{down}}[-\Delta z_i]^+,}
\end{equation}
\textcolor{red}{where $[x]^+ = \max(x, 0)$ is the standard convex rectification function, and $w_{\text{up}} > w_{\text{down}}$ are weight coefficients penalizing altitude gain more heavily than altitude loss. The total convexified energy cost is the sum of the horizontal and vertical terms.}

Regarding the UAV's motion space, according to the original optimization problem, the UAV flight must satisfy the constraints \eqref{1} \eqref{3} \eqref{4}. Except for the collision avoidance constraint, all others are linear or second-order cone constraints. The collision avoidance constraint ensures that the horizontal distance between UAVs remains at least $d_{\text{min}}$. The feasible region for a UAV in a single time slot is the intersection of the maximum horizontal flight distance circle, the environment boundary, and the exterior of circles centered at other UAVs with radius $d_{\text{min}}$. This region may be non-convex due to non-empty intersections between the flight distance circle and collision avoidance circles, as shown in the left part of Fig.~\ref{2}, where the green region represents the feasible domain.

To simplify the problem, a stricter tangent constraint is introduced to guarantee the minimum safety distance. As shown in the right subfigure of Fig.~\ref{2}, by drawing a tangent line at the intersection point between the line connecting the UAVs and the collision avoidance circle, the non-convex feasible region is transformed into a convex feasible region formed by the intersection of the red tangent line and the green circle. Since the primary concern is the horizontal safety distance, the collision avoidance constraint ignores altitude, making the constraint conservative, while altitude control is still incorporated in the motion planning.

In a two-dimensional Cartesian coordinate system, let the current UAV be at \((x_1, y_1)\) and another UAV at \((x_2, y_2)\), with \((x_2, y_2)\) located above and to the left. The equation of the line segment between them is:

\begin{equation}
	y = \frac{y_1 - y_2}{x_1 - x_2} x + \frac{x_1 y_2 - x_2 y_1}{x_1 - x_2}, \quad x_1\neq x_2
\end{equation}
where $x_2 \leq x \leq x_1$ and $y_1 \leq y \leq y_2$. The tangent constraint line passing through the point $(x_2, y_2)$ at a distance of $d_{\min}$ from the line segment can be expressed as:

\begin{equation}
	a x + b y + c \leq 0,
\end{equation}
where $a$, $b$, and $c$ represent the coefficients of the tangent line. For this constraint, the following inequality holds:

\begin{equation} \label{36}
	a x_1 + b y_1 + c \leq 0,
\end{equation}
which indicates that the tangent line's constraint direction is on the side where the current UAV is located, away from the other UAV. Note that if all UAVs are initially deployed such that the minimum horizontal distance constraint is satisfied, and the planned trajectories strictly adhere to the tangent constraints at each time slot, then the current UAV will always remain on the safe side of the tangent line, thus maintaining a safe distance from other UAVs.

Under the condition that the UAV remains within the collision avoidance constraint circle, the following inequality must be satisfied simultaneously except inequality \eqref{36}:

\begin{equation}
	a x_2 + b y_2 + c \geq 0.
\end{equation}

For all other UAVs, each tangent constraint line is obtained, resulting in a total of $U-1$ tangent constraints. The feasible movement region for the UAV within a time slot thus forms a convex set.

\subsubsection{Sub-Optimization Objectives and Problem Formulation}
Considering the consistency between the sub-objectives and the overall optimization objective, the UAV's flight trajectory must balance coverage quality, communication path loss, and energy consumption. Since poor coverage reduces the available UAV computing resources and flying at excessively high altitudes increases the path loss, the sub-objectives need to jointly consider both coverage and the UAV's flight altitude. Furthermore, the UAVs' continuous movement requires significant energy, making energy minimization a crucial component of trajectory planning. 

Based on the above analysis, for a single UAV \(i\), the initial theoretical optimization objective can be constructed as balancing coverage and altitude:
\begin{equation} \label{38}
	\sum_{j=1}^m -s_j^l + \omega_h h_i ,
\end{equation}
where:
\begin{equation} \label{39}
	s_j^l = 
	\begin{cases}
		1, & \sqrt{(x_j - x_i)^2 + (y_j - y_i)^2} \leq z_i \tan(\theta_{\max}) \\
		0, & \sqrt{(x_j - x_i)^2 + (y_j - y_i)^2} > z_i \tan(\theta_{\max})
	\end{cases},
\end{equation}
where \(x_j, y_j\) denote the position coordinates of vehicle \(j\), and \(x_i, y_i, z_i\) denote the 3D coordinates of UAV $i$. \(\theta_{\max}\) is the maximum beam angle, and thus $z_i \tan(\theta_{\max})$ represents the effective coverage radius. \(s_j^l\) is an integer variable indicating whether vehicle \(j\) is within the UAV’s coverage area. Due to the integer variables, this constitutes an integer programming problem. By applying linear relaxation, \(s_j^l\) can be relaxed into a continuous variable $s_j \in[0, 1]$. To formulate a tractable convex problem and directly optimize the altitude $z_i$, we convert the coverage condition into a Second-Order Cone (SOC) constraint:
\begin{equation} \label{40}
	\sqrt{(x_j - x_i)^2 + (y_j - y_i)^2} \leq z_i \tan(\theta_{\max}) + M(1 - s_j),  
\end{equation}
where $0 \leq s_j \leq 1$, and \(M\) represents a sufficiently large constant to ensure the constraint holds when $s_j = 0$. 

However, merely relaxing \(s_j\) cannot precisely reflect the UAV coverage degree, as the solver might indiscriminately assign fractional values to $s_j$. To avoid this, we introduce a soft penalty variable \(N_j\), replace \(M\) with a smaller localized bound, and incorporate the vehicle's instantaneous load \(D_j\) as a weight. 

\textcolor{red}{Moreover, to address the energy consumption, we introduce the energy cost function $J_{\text{eng}, i}$ by incorporating the convexified power models derived in the previous subsection. Specifically, combining the quadratic horizontal energy approximation from \eqref{eq:taylor_ph} and the piecewise linear vertical energy from \eqref{eq:convex_pv}, the energy cost function is defined as:}
\begin{equation} \label{eq:energy_term}
	\begin{split}
		\textcolor{red}{J_{\text{eng}, i}} &\textcolor{red}{= w_{\text{eng}} \Big( c_{xy} \underbrace{\| \mathbf{p}_i - \mathbf{p}_{i}^{\mathrm{p}} \|_2^2}_{\text{from Eq.~\eqref{eq:taylor_ph}}} } \\
		&\textcolor{red}{+ \underbrace{c_{\text{up}}[z_i - z_{i}^{\mathrm{p}}]^+ + c_{\text{down}}[-(z_i - z_{i}^{\mathrm{p}})]^+}_{\text{from Eq.~\eqref{eq:convex_pv}}} \Big),}
	\end{split}
\end{equation}
\textcolor{red}{where the first term represents the horizontal flight energy cost proportional to the squared displacement, and the subsequent terms capture the asymmetric vertical energy cost. Here, $\mathbf{p}_i = (x_i, y_i)$ and $[x]^+ = \max(x, 0)$ is the convex rectification function.}

\textcolor{red}{Considering the difficulty of simultaneously optimizing multiple UAV positions with collision avoidance constraints, we adopt a sequential distributed optimization method. By limiting the sensing range $s_r$, the UAV focuses on the local vehicle distribution. For UAV \(i\), the comprehensive convex optimization problem is expressed as follows:}
\begin{subequations}\label{eq:local_opt}
	\begin{align}
		\min_{\substack{x_i, y_i, z_i, \\ \mathbf{s}, \mathbf{r}, \mathbf{N}}} \quad & \beta_1 \sum_{j=0}^{n_{i,\text{local}}-1} r_j + \beta_2 \sum_{j=0}^{n_{i,\text{local}}-1} N_{j} + \beta_3 z_i + J_{\text{eng}, i} \label{eq:local_obj} \\
		\text{s.t.} \quad 
		& \text{Constraints \eqref{3}, \eqref{36}}, \label{eq:local_con_collision} \\
		& \| \mathbf{p}_{j} - \mathbf{p}_i \|_2 \leq z_i \tan(\theta_{\max}) + M_{\text{linear}} (1 - s_j) + N_{j}, \label{eq:local_con_soc} \\
		& s_{j}(2D_j)^2 \geq -r_j, \label{eq:local_con_load} \\
		& \| \mathbf{p}_i - \mathbf{p}_{i}^{\mathrm{p}} \|_2^2 \leq (l_{\max}^{h})^2, \label{eq:local_con_move_h} \\
		& x_{u}^{\min} \leq x_i \leq x_{u}^{\max}, \quad y_{u}^{\min} \leq y_i \leq y_{u}^{\max}, \label{eq:local_con_bound_xy} \\
		& z_{u}^{\min} \leq z_i \leq z_{u}^{\max}, \label{eq:local_con_bound_z} \\
		& z_{i}^{\mathrm{p}} - l_{\max}^{v} \leq z_i \leq z_{i}^{\mathrm{p}} + l_{\max}^{v}, \label{eq:local_con_move_v} \\
		& 0 \leq s_j \leq 1, \quad r_{\min} \leq r_j \leq 0, \quad 0 \leq N_{j} \leq N_{\max}, \label{eq:local_con_vars}
	\end{align}
\end{subequations}
where $n_{i,\text{local}}$ denotes the number of vehicles within the sensing range of UAV \(i\). \textcolor{red}{The vectors $\mathbf{s}, \mathbf{r}, \mathbf{N}$ represent the sets of auxiliary variables $\{s_j\}, \{r_j\}, \{N_j\}$ introduced for convex relaxation.} The optimization objective \eqref{eq:local_obj} is a weighted linear combination of load-based coverage slack $r_j$, distance penalty $N_j$, altitude penalty $z_i$, and the energy cost $J_{\text{eng}, i}$. $\beta_1$, $\beta_2$, and $\beta_3$ represent the corresponding weights. 

\textcolor{red}{Constraints \eqref{eq:local_con_collision} enforce the minimum separation distance and the tangent-based collision avoidance rules derived in the previous section.} Eq.~\eqref{eq:local_con_soc} defines the SOC coverage constraint directly determining the objective values. When a vehicle is inside the UAV’s coverage, its distance to the UAV is less than \(z_i \tan(\theta_{\max})\), which enforces \(N_j = 0\) and \(s_{j} = 1\). According to Eq.~\eqref{eq:local_con_load}, as \(s_{j}\) increases, the slack variable \(r_j\) can take smaller (more negative) values. Reflected in the objective function, smaller values of \(N_j\) and \(r_j\) are preferred. When the distance exceeds the coverage radius, a trade-off must be made: the solver balances whether to move the UAV (incurring energy cost), increase altitude $z_i$ (increasing path loss penalty), or increase $N_j$ and decrease $s_j$ (abandoning coverage). 
Constraints \eqref{eq:local_con_move_h} and \eqref{eq:local_con_move_v} limit the UAV’s horizontal and vertical movement distances per time slot. Constraints \eqref{eq:local_con_bound_xy} and \eqref{eq:local_con_bound_z} define the boundaries of the 3D environment. Eq.~\eqref{eq:local_con_vars} provides the feasible ranges for the auxiliary variables.

Since all objective terms are convex and all constraints are linear or SOC forms, this is a standard Second-Order Cone Programming (SOCP) problem. \textcolor{red}{It can be directly and efficiently solved using open-source interior-point solvers. To extend this to multi-UAV coordination without relying on a centralized controller, we propose a distributed sequential optimization framework. Specifically, each UAV independently solves its local SOCP problem in a predefined sequence. Upon determining its optimal trajectory, the UAV broadcasts its updated 3D position and the IDs of its covered vehicles via lightweight inter-UAV communication. To prevent redundant coverage and load imbalance, subsequent UAVs employ a cooperative "masking" mechanism, deliberately excluding the already-covered vehicles from their local objective functions. This proactive distributed strategy strictly ensures collision avoidance while inherently steering the swarm to maximize the global service area.}
\textcolor{red}{\subsection{Resource Scheduling Based on DRL and LLM}}

\textcolor{red}{Given fixed UAV and vehicle positions, communication link states are determined. However, the strong coupling between transmit power and RB occupancy renders the joint resource and task allocation NP-hard. While task allocation is efficiently solvable via Linear Programming, the high-dimensional hybrid discrete-continuous action space of resource scheduling poses significant challenges to traditional DRL, often leading to performance degradation in unexplored scenarios. To address this, we propose a two-stage mechanism: utilizing DRL for global centralized initial scheduling, followed by LLM-based task reallocation based on execution outcomes.}

\begin{figure*}[t]
	\centering
	\includegraphics[trim=0.5cm 0.5cm 0.5cm 0.5cm, clip, width=0.85\textwidth]{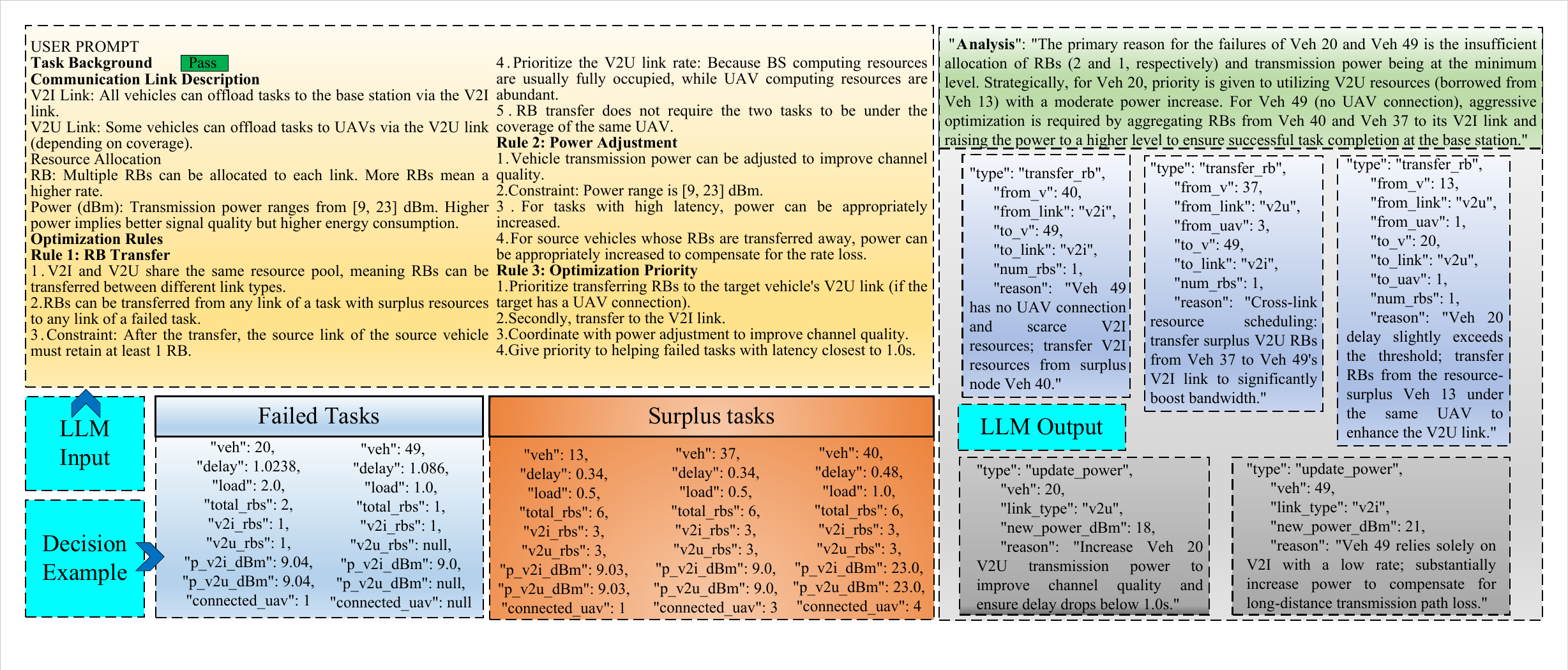}
	\caption{LLM Prompts and Decision Examples}
	\label{LLM}
\end{figure*}
\subsubsection{\textcolor{red}{DRL-Based Centralized Initial Resource Scheduling}}

\textcolor{red}{We formulate the joint resource block and power allocation problem as a Markov Decision Process (MDP). Given the continuous nature of the action space, the Deep Deterministic Policy Gradient (DDPG) algorithm is employed. At each time step $t$, the central agent observes the environment state—incorporating UAV trajectory and task offloading decisions—and outputs scheduling actions. The specific definitions of the MDP tuples are as follows:}

\begin{itemize}
	\item \textcolor{red}{\textbf{State Space:} The state captures the network load and channel conditions essential for decision-making. For each vehicle $i \in \{1, \dots, M\}$, the local normalized observation $s_i$ consists of the task load $D_i$, V2I/V2U channel qualities ($Q_{v2i}^i, Q_{v2u}^i$), and the UAV connection indicator $z_1^i \in \{0,1\}$. The local state $s_i$ and the global state vector $\mathcal{S}$ are defined as:}
	\textcolor{red}{
	\begin{equation}
		s_i = \{D_i, Q_{v2i}^i, Q_{v2u}^i, z_1^i\},  \mathcal{S} = \{s_1, s_2, \dots, s_{M}\}
	\end{equation}}
	\vspace{-0.3cm}
	\item \textcolor{red}{\textbf{Action Space:} Direct allocation of discrete RBs leads to dimensionality explosion. To mitigate this, we decompose the action into continuous power control and resource priority. For vehicle $i$, the action is defined as $a_i =[P_{norm}^i, r_f^i]$, where $P_{norm}^i \in (0, 1)$ denotes the normalized transmit power and $r_f^i \in (0, 1)$ represents the scheduling priority. The global action vector $\mathcal{A}$ maintains a fixed dimension of $M \times 2$. During execution, the environment calculates the RB quota for each vehicle based on the weighted proportion of priorities $r_f$, followed by a greedy assignment of specific RB indices based on optimal channel gains.}
	
	\item \textcolor{red}{\textbf{Reward Function:} To align the learning objective with system optimization, the reward is derived from the LP solution. Upon action execution and resource mapping, the system calculates the current objective function value. The immediate reward is defined as the negative of this cost to guide the agent toward minimization:}
	\textcolor{red}{
		\begin{equation}
			r_t = -\left( \omega_1 T + \omega_2 E + \omega_3 \sum_{m=1}^{M} \xi_m\right).
	\end{equation}}
\end{itemize}
\vspace{-0.4cm}

\textcolor{red}{To handle high-dimensional states and continuous action spaces, we design a DDPG-based Actor-Critic architecture utilizing the following neural network structures:}

\begin{itemize}
	\item \textcolor{red}{\textbf{Actor Network:} The policy network $\mu(s|\theta^\mu)$ employs a Multi-Layer Perceptron (MLP). To prevent gradient vanishing and accelerate feature extraction, Layer Normalization is applied to the hidden layers. A Sigmoid activation function is used at the output layer to strictly bound the power control factors and priority weights within the $(0, 1)$ interval.}
	
	\item \textcolor{red}{\textbf{Critic Network:} The value network $Q(s, a|\theta^Q)$ adopts a dual-stream architecture. State features $\mathcal{S}$ and action features $\mathcal{A}$ are extracted via independent linear layers before being concatenated in deep hidden layers. This fused representation is mapped to a scalar Q-value, estimating the expected long-term return.}
\end{itemize}

\textcolor{red}{During training, Gaussian noise is added to the Actor's output to encourage exploration of the action space. Transitions $(s_t, a_t, r_t, s_{t+1})$ are stored in a Replay Buffer. To ensure training stability and break temporal correlations between sequential data, the networks are updated offline using random mini-batches of size $B_{batch}$ sampled from the buffer.}

\textcolor{red}{The update of the Critic network is based on the Bellman optimality equation. For each sampled transition $i$, the target Q-value $y_i$ is composed of the immediate reward from the environment and the discounted Q-value of the next state predicted by the target network:}
\textcolor{red}{
	\begin{equation}
		y_i = r_i + \gamma Q'(s_{i+1}, \mu'(s_{i+1}|\theta^{\mu'}) | \theta^{Q'})
\end{equation}}
\textcolor{red}{where $\gamma$ is the discount factor, and $\mu'$ and $Q'$ represent the target Actor and target Critic networks, respectively. The Critic network calculates the loss function by minimizing the Mean Squared Error (MSE) between the online network prediction and the target value:}
\textcolor{red}{
	\begin{equation}
		L(\theta^Q) = \frac{1}{B_{batch}} \sum_{i=1}^{B_batch} \left( y_i - Q(s_i, a_i | \theta^Q) \right)^2
\end{equation}}
\textcolor{red}{Subsequently, the parameters $\theta^Q$ of the Critic network are updated using the gradient descent method.}

\textcolor{red}{The update of the Actor network relies on the Deterministic Policy Gradient theorem. Its optimization objective is to find the policy parameters that maximize the value evaluated by the Critic network. Specifically, the algorithm fixes the parameters of the Critic network, uses the output of the Actor network as the action input to the Critic, and calculates the gradient of the Q-value with respect to the Actor network parameters $\theta^\mu$ via the chain rule:}
\textcolor{red}{
	\begin{equation}
		\nabla_{\theta^\mu} J \approx \frac{1}{M} \sum_{i=1}^{M} \nabla_a Q(s, a|\theta^Q)|_{s=s_i, a=\mu(s_i)} \cdot \nabla_{\theta^\mu} \mu(s|\theta^\mu)|_{s=s_i}
\end{equation}}
\textcolor{red}{The Actor network is then updated using the gradient ascent method.}

\textcolor{red}{Finally, to ensure training stability, the target networks do not directly copy the parameters of the online networks but instead employ a Soft Update strategy. After each iteration, the target network parameters slowly track the changes in the online network parameters with a very small learning rate $\tau$:}
\textcolor{red}{
	\begin{equation}
		\theta^{Q'} \leftarrow \tau \theta^Q + (1 - \tau) \theta^{Q'}
\end{equation}}
\textcolor{red}{
	\begin{equation}
		\theta^{\mu'} \leftarrow \tau \theta^\mu + (1 - \tau) \theta^{\mu'}
\end{equation}}
\textcolor{red}{Through the aforementioned interaction and alternating update process, the DRL agent gradually converges to an approximately optimal resource scheduling policy, thereby providing a high-quality initial allocation baseline for outlier assessment and long-tail task reallocation in the subsequent LLM stage.}

\subsubsection{\textcolor{red}{LLM-Based Macro-Adjustment Method}}

\textcolor{red}{As discussed, while DRL ensures efficient decision-making in typical scenarios, its limited exploration often causes severe performance degradation in long-tail or unexplored environments due to poor generalization. To overcome this inherent limitation, we integrate LLMs into our framework. Pre-trained on massive datasets, LLMs possess strong logical reasoning and generalization capabilities. Leveraged by In-Context Learning (ICL), LLMs can execute zero-shot or few-shot reasoning, demonstrating robust adaptability to novel tasks without retraining \cite{zhang2025}. Furthermore, Chain-of-Thought (CoT) prompting enables transparent, human-like inference, providing strong interpretability for the scheduling decisions. Consequently, we propose an LLM-based macro-adjustment framework to monitor and refine DRL-generated policies, effectively alleviating the generalization bottleneck.}

\textcolor{red}{Within the alternating optimization framework, given the fixed UAV and vehicle positions, the DRL agent first determines the initial RB and transmit power allocations, followed by the LP-based task offloading scheme. Based on this initial allocation, the system estimates the task completion time for each vehicle to verify compliance with strict latency constraints. This evaluation rapidly identifies two extreme subsets: \textit{failed tasks} (violating the latency threshold) and \textit{surplus tasks} (completing well ahead of schedule with abundant redundant resources). Targeting these edge cases, the LLM acts as a centralized macro-scheduler to reallocate communication resources and fine-tune transmit powers. This semantic intervention mitigates resource imbalances and rescues failing tasks, ultimately maximizing the overall task success rate.}

\textcolor{red}{Despite the massive parameter scale of typical LLMs, deploying quantized lightweight models at the network edge has become increasingly viable. Furthermore, because our scheduling task relies on highly structured prompts, KV caching can be fully leveraged. By caching the attention keys and values of static prompt prefixes, the system eliminates redundant forward-pass computations during continuous invocations, thereby drastically reducing inference latency and computational overhead at the edge.}

\textcolor{red}{To instantiate this framework, a customized prompt template is designed to encode the specific networking constraints. Additionally, advanced large-scale LLMs are utilized offline to generate high-quality scheduling examples, providing few-shot guidance for the deployed edge model. Consequently, the input sequence is structured into three distinct components: the system prompt ($\mathcal{P}_{sys}$), the few-shot examples ($\mathcal{P}_{eg}$), and the dynamic observation data ($\mathcal{P}_{data}$). This structured input is formally expressed as the concatenated token sequence $\mathcal{X}$:}
\textcolor{red}{
	\begin{equation}
		\mathcal{X} = \left[ \mathcal{P}_{sys} \oplus \mathcal{P}_{eg} \oplus \mathcal{P}_{data} \right]
\end{equation}}
\textcolor{red}{where $\oplus$ denotes the token concatenation operation. The prompt instructions and examples ($\mathcal{P}_{sys} \oplus \mathcal{P}_{eg}$) typically occupy the vast majority of the context window and remain strictly static. Consequently, their KV cache is pre-computed and stored. The dynamic data portion $\mathcal{P}_{data}$, formatted as serialized JSON arrays detailing the failed and surplus tasks, constitutes only a minimal fraction of the tokens, meaning the LLM only needs to compute the attention weights for this newly injected data.}

\textcolor{red}{Upon receiving the input sequence $\mathcal{X}$, the LLM conducts contextual analysis and generates a set of adjustment actions alongside a brief reasoning analysis. Based on the system constraints, the LLM is authorized to execute two types of macro-actions: RB Transfer and Power Update. For RB transfer, the LLM can intelligently shift resources across different links (e.g., from a surplus V2I link to a failing V2U link), under the hard constraint that the source link must retain at least one RB. For power updates, the transmit power of critically delayed links can be boosted within the feasible bounds $[P_{\min}, P_{\max}]$ to compensate for poor channel qualities. The output generation process is formulated as:}
\textcolor{red}{
	\begin{equation}
		\mathcal{A}_{LLM}, \mathcal{R}_{CoT} \sim \text{LLM}(\mathcal{X})
\end{equation}}
\textcolor{red}{where $\mathcal{A}_{LLM}$ represents the parsed, strictly JSON-formatted list of optimal actions (e.g., \texttt{transfer\_rb}, \texttt{update\_power}), and $\mathcal{R}_{CoT}$ denotes the natural language rationale explaining the underlying optimization logic. Through this LLM-guided macro-adjustment, the system effectively compensates for the heuristic flaws of the DRL agent in edge-case scenarios, achieving a highly reliable and interpretable resource scheduling paradigm.}

\textcolor{red}{Fig.~\ref{LLM} presents the key prompts provided to the LLM alongside a concrete decision-making example. Given the length of the complete prompt, the figure highlights the core decision rules; detailed definitions of actions, the semantics of input data, and the decision background are available in the code repository linked earlier. As observed in the example, the LLM not only clearly specifies the actions to be executed but also articulates the rationales behind them. This significantly enhances the interpretability of the decision-making process, highlighting a unique advantage of utilizing LLMs.}
\subsection{\textcolor{red}{LP-Based Task Offloading Method}}

\textcolor{red}{Once the UAV trajectory and initial resource allocation are determined, the communication link states are fixed. Consequently, the optimization variables are reduced to task offloading ratios and queuing delays. However, according to equations \eqref{21} and \eqref{22}, queuing delay depends on the specific data arrival order, which is strictly coupled with offloading decisions. This coupling renders the queuing delay incapable of being expressed in a closed form, making the original problem non-convex.}

\textcolor{red}{To decouple these variables, we incorporate task offloading into an alternating optimization framework. We introduce an estimated queuing delay parameter $\hat{Q}_{m}^{k}$ for vehicle $m$ at node $k$ (where $k \in \{I, u\}$ represents the BS or UAV $u$). To ensure accurate estimation throughout the optimization process, $\hat{Q}_{m}^{k}$ is defined as a piecewise function of the iteration index $i$:}
\textcolor{red}{
	\begin{equation}
		\label{eq:queuing_est}
		\hat{Q}_{m}^{k} = 
		\begin{cases} 
			\bar{Q}^k(D_m), & \text{if } i = 0 \\ 
			Q_{m, \text{prev}}^{k}, & \text{if } i > 0 
		\end{cases}
\end{equation}}
\textcolor{red}{where $Q_{m, \text{prev}}^{k}$ denotes the actual queuing delay calculated from the solution of the previous iteration. For the initialization phase ($i=0$), we employ a load-aware historical average. Let $\mathcal{Q}_{hist}^k(D)$ denote the set of historical queuing delays at node $k$ under a specific task load $D$. The average queuing delay $\bar{Q}^k(D_m)$ corresponds to the current task load $D_m$ of vehicle $m$, calculated as:}
\textcolor{red}{
	\begin{equation}
		\bar{Q}^k(D_m) = \frac{1}{|\mathcal{Q}_{hist}^k(D_m)|} \sum_{q \in \mathcal{Q}_{hist}^k(D_m)} q,
\end{equation}}
\textcolor{red}{where $|\cdot|$ denotes the cardinality. During the iterative process ($i > 0$), since the macroscopic load distribution at edge nodes remains relatively stable during LLM fine-tuning, the queuing delay from the previous LP solution serves as a valid estimate.}

\textcolor{red}{By fixing the queuing delays to the estimated values $\hat{Q}_{m}^{I}$ and $\hat{Q}_{m}^{u}$, the overall optimization problem simplifies to a standard LP problem with respect to the offloading ratios $\boldsymbol{\gamma}$:}
\begin{subequations}\label{eq:global_opt}
	\setlength{\jot}{2pt}
	\begin{align}
		\min_{\boldsymbol{\gamma}} \quad & \omega_1 T_m + \omega_2 E + \omega_3 \sum_{m=1}^{M} \xi_m, \label{eq:obj} \\
		\text{s.t.} \quad
		& \gamma_{m}^o + \gamma_{m}^I + \sum_{u=1}^{U} \alpha_{m}^u \gamma_{m}^u = 1, \ \forall m, \label{eq:con_alloc} \\
		& \sum_{m=1}^{M} \alpha_{m}^u D_m \gamma_{m}^u \leq D_{\max}^u, \ \forall u, \label{eq:con_uav_cap} \\
		& \sum_{m=1}^{M} D_m \gamma_{m}^I \leq D_{\max}^I, \label{eq:con_inf_cap} \\
		& T_m \geq \frac{D_m c}{f_m} \gamma_{m}^o, \ \forall m, \label{eq:con_local_time} \\
		& T_m \geq \left( \frac{D_m}{R_{m}^I} + \frac{D_m c}{f_I} \right) \gamma_{m}^I + \hat{Q}_{m}^I, \ \forall m, \label{eq:con_inf_time} \\
		& T_m \geq \sum_{u=1}^{U} \left( \frac{D_m}{R_{m}^u} + \frac{D_m c}{f_u} \right) \gamma_{m}^u + \hat{Q}_{m}^u, \ \forall m, \label{eq:con_uav_time} \\
		& T_m \leq T_{m}^{\max} + \xi_m, \ \forall m, \label{eq:con_deadline} \\
		& T_m, \xi_m \geq 0; \ \gamma_{m}^o, \gamma_{m}^I, \gamma_{m}^u \in [0,1]. \label{eq:con_var_bounds}
	\end{align}
\end{subequations}
\textcolor{red}{Where constraints \eqref{eq:con_uav_cap}--\eqref{eq:con_inf_cap} define the capacity limits. Constraint \eqref{eq:con_deadline} imposes latency thresholds with a slack variable $\xi_m$. Crucially, in constraints \eqref{eq:con_inf_time} and \eqref{eq:con_uav_time}, $\hat{Q}_{m}^I$ and $\hat{Q}_{m}^u$ are constant parameters determined by Eq. \eqref{eq:queuing_est}, transforming the delay calculation into a linear Min-Max framework. Consequently, the problem can be efficiently solved using standard LP algorithms.}

\textcolor{red}{Algorithm \ref{alg:overall_framework} delineates the hierarchical execution flow of the proposed joint optimization framework. In each decision slot, the process commences with distributed trajectory planning, where UAV positions are updated via sequential convex optimization to adapt to the dynamic vehicle topology. Subsequently, the resource scheduling enters a closed-loop ``DRL-LP-LLM-LP'' sequence. The DRL agent first generates provisional resource and power allocations, serving as the basis for a preliminary LP solution. This step serves a dual purpose: quantifying the intrinsic cost $\Psi_{drl}$ of the DRL policy and identifying long-tail failure tasks via estimated completion times. Consequently, the LLM performs semantic macro-adjustments on the resource allocation, and the refined state drives a second LP solve to derive the final offloading ratios $\boldsymbol{\gamma}^*$. Crucially, during the network update phase, the reward signal stored in the replay buffer is strictly coupled with the DRL's original action. This reward decoupling mechanism effectively prevents policy gradient bias caused by external LLM interventions, ensuring the stability and convergence of the reinforcement learning process.}
\begin{algorithm}[t]
	\small
	\caption{Joint Trajectory Control and Resource Scheduling based on DRL and LLM}
	\label{alg:overall_framework}
	\KwIn{Network topology, time slots $K$, vehicles $M$, UAVs $U$, LLM prompts $\mathcal{P}_{sys}$ and $\mathcal{P}_{eg}$.}
	\KwOut{Sequences of optimal $\mathbf{T}^*$, $\mathbf{R}^*$, $\mathbf{P}^*$, and $\boldsymbol{\gamma}^*$.}
	
	Initialize DDPG networks $\theta^\mu, \theta^Q, \theta^{\mu'}, \theta^{Q'}$, replay buffer $\mathcal{D}$\;
	Pre-compute static LLM KV cache: $\mathcal{C}_{KV} \gets \text{KV Cache}(\mathcal{P}_{sys} \oplus \mathcal{P}_{eg})$\;
	Initialize historical queuing delay estimates $\hat{\mathbf{Q}}$\;
	
	\For{$t = 1$ \KwTo $K$}{
		\tcp{1. Sequential UAV Optimization}
		For u = 1 to U, sequentially update position $(x_u(t), y_u(t), z_u(t)) \in C_u$\;
		Update communication link states and channel qualities\;
		
		\tcp{2. DRL-based Initial Scheduling}
		Observe state $s_t$, generate action $a_t = \mu(s_t | \theta^\mu) + \mathcal{N}$\;
		Map $a_t$ to initial resource allocation $\mathbf{R}_{drl}$ and power $\mathbf{P}_{drl}$\;
		
		\tcc{Solve 1st LP to estimate times and extract delays for next steps}
		Solve LP \eqref{eq:global_opt} given $(\mathbf{R}_{drl}, \mathbf{P}_{drl}, \hat{\mathbf{Q}})$ to obtain cost $\Psi_{drl}$, task completion times $\mathbf{T}_{comp}$, and updated queuing delays $\mathbf{Q}_{new}$\;
		Update $\hat{\mathbf{Q}} \gets \mathbf{Q}_{new}$ for the subsequent LP optimization\;
		
		\tcp{3. LLM-based Macro-Adjustment}
		Identify \texttt{failed\_tasks} and \texttt{surplus\_tasks} based on $\mathbf{T}_{comp}$\;
		\uIf{\texttt{failed\_tasks} $\neq \emptyset$}{
			Construct dynamic data prompt $\mathcal{P}_{data}$ and full sequence $\mathcal{X} \gets[\mathcal{P}_{sys} \oplus \mathcal{P}_{eg} \oplus \mathcal{P}_{data}]$\;
			Obtain macro-actions $\mathcal{A}_{LLM} \sim \text{LLM}(\mathcal{X})$ leveraging $\mathcal{C}_{KV}$\;
			Apply valid actions after constraint checking to obtain $\mathbf{R}_{llm}, \mathbf{P}_{llm}$\;
		}
		\Else{
			Retain initial allocations: $\mathbf{R}_{llm} \gets \mathbf{R}_{drl}$, $\mathbf{P}_{llm} \gets \mathbf{P}_{drl}$\;
		}
		
		\tcc{Solve 2nd LP using LLM-adjusted resources and updated delays}
		Solve LP \eqref{eq:global_opt} given $(\mathbf{R}_{llm}, \mathbf{P}_{llm}, \hat{\mathbf{Q}})$ to obtain final $\boldsymbol{\gamma}^*_t$, actual cost $\Psi_t^*$, and final delays $\mathbf{Q}^*$\;
		Update $\hat{\mathbf{Q}} \gets \mathbf{Q}^*$ for the next time slot $t+1$\;
		
		\tcp{4. DDPG Network Update}
		\textbf{Store tuple} $(s_t, a_t, r_t = -\Psi_{drl}, s_{t+1})$ \textbf{into} $\mathcal{D}$ 
		\If{$|\mathcal{D}| \geq B_{batch}$}{
			Sample mini-batch of size $B_{batch}$ from $\mathcal{D}$, compute target $y_i$\;
			Update Critic $\theta^Q$ by minimizing MSE loss $L(\theta^Q)$\;
			Update Actor $\theta^\mu$ by maximizing policy gradient $\nabla_{\theta^\mu} J$\;
			Soft update target networks $\theta^{Q'}$ and $\theta^{\mu'}$\;
		}
	}
	\Return Sequence of $(\mathbf{T}^*, \mathbf{R}^*, \mathbf{P}^*, \boldsymbol{\gamma}^*)$\;
\end{algorithm}
\vspace{-0.2cm}
\subsection{\textcolor{red}{Convergence and Complexity Analysis}}
\textcolor{red}{The proposed joint optimization framework is decoupled into two sequential stages: Block Coordinate Descent (BCD)-based UAV trajectory planning, and a hybrid DRL-LLM-LP closed-loop for resource and task scheduling. In the first stage, the multi-UAV cooperative trajectory problem is solved by sequentially optimizing each UAV's trajectory while fixing the others, formulating each sub-problem as a SOCP. Since the objective function is lower-bounded within a compact feasible region and each SOCP is strictly convex, standard convex analysis guarantees that this sequential update monotonically converges to a local optimum. }

\textcolor{red}{Subsequently, with the trajectories fixed, the system minimizes the total cost $\mathcal{J}$, which encompasses latency, energy, and penalties, through the iterative DRL $\to$ LP $\to$ LLM $\to$ LP execution loop. Although DRL and LLMs are fundamentally heuristic, the embedded LP guarantees the globally optimal offloading ratios $\boldsymbol{\gamma}^*$ for any intermediate resource allocation state. Furthermore, the LLM is explicitly prompted to minimize the penalty $\xi_m$ of long-tail tasks and only intervenes when performance bottlenecks are detected. This conditional semantic intervention probabilistically ensures the monotonic non-increase of $\mathcal{J}$. Given the lower-bounded nature of the physical constraints, the overall algorithm reliably converges to a stable solution within finite iterations.}

\textcolor{red}{The total computational overhead is decomposed into three components: trajectory planning, neural network inference, and linear programming. Let $L_{traj}$ and $L_{loop}$ denote the number of iterations for the trajectory and resource scheduling stages, respectively. First, solving the trajectory SOCP via the Interior Point Method (IPM) for a single UAV involves $O(K)$ variables and $O(KU)$ constraints, yielding a complexity of $\mathcal{O}(K^{3.5})$. Accounting for the vehicle-load-based objective construction, the total trajectory planning complexity scales as $\mathcal{O}(L_{traj} U (K^{3.5} + MK))$. Second, the DRL Actor network incurs a negligible inference overhead of $\mathcal{O}(\sum n_l n_{l-1})$. For the LLM, adopting a Mixture-of-Experts (MoE) architecture (e.g., Qwen3-235B-A22B) significantly curtails the computational burden by activating only $P_{act} \approx 22 \times 10^9$ parameters per token. Furthermore, by leveraging KV caching to pre-compute static prompt attention, the inference complexity scales linearly only with the dynamic input and output tokens, culminating in $\mathcal{O}((S_{dyn} + S_{out}) P_{act})$. Finally, solving the $3M$-variable task offloading LP via IPM requires $\mathcal{O}(M^{3.5})$ operations. Executed twice per scheduling loop, this contributes $\mathcal{O}(M^{3.5})$ per loop. In summary, the total computational complexity per decision slot is formulated as:}
\textcolor{red}{
\begin{equation}
	\begin{split}
		C_{total} &\approx O\bigg( L_{traj} U (K^{3.5} + MK) + \\
		&\quad L_{loop} \left[ (S_{dyn} + S_{out}) P_{act} + 2 M^{3.5} \right] \bigg)
	\end{split}
\end{equation}}
\textcolor{red}{Given that $M$, $U$, and $K$ are finite constants and the MoE architecture significantly mitigates the computational burden of the LLM, the proposed algorithm demonstrates feasibility for deployment at the network edge while ensuring superior system performance.}
\begin{table}[!t]
	\centering
	\caption{System Parameters}
	\label{tab:parameters}
	\small 
	\begin{tabular*}{\columnwidth}{@{\extracolsep{\fill}} p{2.5cm} p{1.2cm} | p{2.5cm} p{1.2cm}}
		\toprule
		\textbf{Parameter} & \textbf{Value} & \textbf{Parameter} & \textbf{Value} \\
		\midrule
		Number of vehicles & 50 & Carrier frequency & 2.4 GHz \\
		Number of UAVs & 5 & Parameter $\omega_a$ & 9.61 \cite{23}\\
		Slot duration & 1 s & Parameter $\omega_b$ & 0.16 \cite{23}\\
		UAV max horizontal movement & 15 m & BS computation frequency & 9 GHz \\
		UAV max vertical movement & 10 m & UAV computation frequency & 5 GHz \\
		LoS angle & 42.44$^\circ$ \cite{10}& OBU frequency & 0.5 GHz \\
		UAV sensing range & 70 m & Delay weight $\omega_1$ & 1 \\
		Min flight altitude & 50 m & Energy weight $\omega_2$ & 0.001/0.02 \\
		Max flight altitude & 100 m & Delay weight $\omega_3$ & 5 \\
		$\eta_{\text{LoS}}$ & 1 dB & Task deadline & 1 s \\
		$\eta_{\text{NLoS}}$ & 20 dB & Total bandwidth $B$ & 10 MHz \\
		RB bandwidth & 180 kHz & CPU cycles per bit & 1000 \\
		Vehicle speed & 10--15 m/s & Vehicle task size & 0/0.5/1/2 Mb \\
		Capacitance activation coefficient & $10^{-27}$ & Total load per slot & 30--50 Mb \\
		\bottomrule
	\end{tabular*}
\end{table}
\begin{figure*}[t]
	\centering
	\includegraphics[trim=0.5cm 3.5cm 1cm 3.5cm, clip, width=0.85\textwidth]{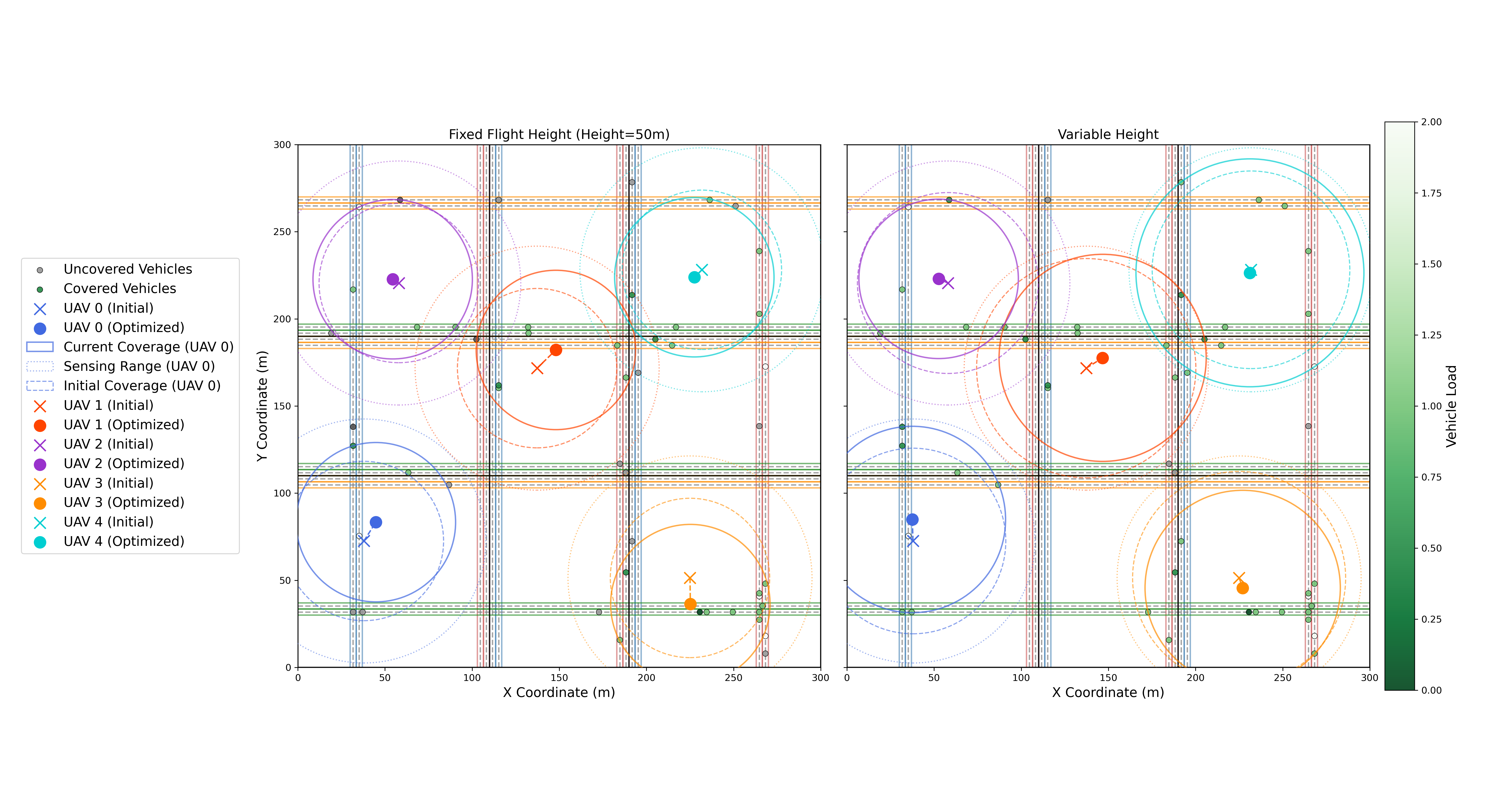}
	\vspace{-0.2cm}
	\caption{Comparison between fixed height (50m) and variable height strategies}
	\label{position_coverage}
\end{figure*}
\section{Simulation Results}\label{simulation results}
In this section, simulation experiments are conducted based on the considered high-density vehicular environment. The simulation platform is implemented in Python, with convex optimization problems solved using the \texttt{cvxpy} and \texttt{linprog} libraries. The LLM employed is the open-source \texttt{Qwen3-235B-A22B}. Numerical and visualization results verify the effectiveness of the proposed method. The main simulation parameters are summarized in \TableRef{tab:parameters}.

\textcolor{red}{In the comparative experiments, UAV flight trajectory control is implemented using Convex Optimization (CVX), a Large Vision Model (LVM), and Multi-Agent Deep Q-Networks (MADQN). The resource allocation component compares our collaborative DDPG \& LLM approach against a standalone DDPG baseline (ablation study). Task ratio allocation is performed using either the MADDPG algorithm or a LP algorithm. Corresponding to the three subproblems addressed in this paper, the combinations of comparison algorithms are configured as follows:}
\begin{itemize}
	\item \textbf{Proposed}: CVX (Trajectory) + DDPG \& LLM (Resource) + LP (Task)
	\item \textbf{DRL-Resource (No LLM)}: CVX + DDPG + LP
	\item \textbf{MADQN-Traj}: MADQN + DDPG + LP
	\item \textbf{LVM-Traj}: LVM + DDPG + LP
	\item \textbf{Full-MADRL}: MADQN + DDPG + MADDPG
	\item \textbf{MADDPG-Task}: CVX + DDPG + MADDPG
\end{itemize}
These combinations are utilized to decouple and evaluate the impact of different modules on overall system performance.

\vspace{-0.5cm}
\subsection{UAV Path Planning Performance}
To evaluate the performance of UAV path planning, in conjunction with Eq.~\eqref{38} and comprehensively considering the number of covered vehicles, UAV flight altitude, and flight energy consumption, the performance metric $R_t$ is defined as:  
\begin{equation}
	R_t = \sum_{j=1}^M s_j^l - \omega_h \sum_{u=1}^U h_u - \omega_e E_{flight},
\end{equation}
where $\omega_h$ and $\omega_e$ are weights for altitude penalty and energy consumption, respectively.

Fig.~\ref{position_coverage} illustrates the UAV positions and coverage under a single decision instance for both fixed and variable UAV altitude strategies. Roads in different directions are represented by distinct colors, and the optimization order follows the UAV indices sequentially. In the left subfigure, it is observed that with a fixed altitude and dispersed vehicle distribution, each UAV can cover only a small fraction of vehicles. Coverage is noticeably improved in the right subfigure, where UAVs tend to shrink their coverage radii to precisely encompass target vehicles, avoiding unnecessarily large coverage areas—as exemplified by UAV 1 (orange), UAV 2 (purple), and UAV 3 (yellow). Moreover, the flexibility afforded by altitude variation enables the algorithm to better identify vehicles near the coverage boundary, thereby optimizing the objective value $R_t$ through altitude adjustments.

\textcolor{red}{Fig.~\ref{fig:uav_trace} depicts the flight trajectories of five UAVs over 20 consecutive time slots, marking the vehicle positions and UAV coverage at the final slot. It can be seen that UAVs tend to cruise within their respective local regions. This is because, when constructing the observation space for each UAV, we masked vehicles already covered by peers, implicitly promoting cooperative coverage among the swarm. Maintaining such a loosely distributed formation maximizes the total coverage area, allowing each UAV to capture local vehicle distribution changes and adjust its position accordingly.}

\textcolor{red}{Fig.~\ref{path_planning_performance} compares the performance of different trajectory planning algorithms under metric $R_t$. Due to fixed initial positions, the first optimization round typically involves significant adjustments in altitude and position, resulting in higher energy consumption and lower metric values; subsequent steps tend to stabilize. The CVX-based method significantly outperforms others, as convex optimization guarantees finding the global optimum. In contrast, LVM-based and MADQN-based methods perform worse, as learning-based approaches often converge to local optima. Under fixed altitude constraints, the performance gap between CVX and learning-based methods narrows, indicating that the decision space in the altitude dimension is critical for performance enhancement.}

\subsection{Performance Comparison}
\begin{figure*}[t]
	\centering
	
	\begin{minipage}{0.28\textwidth}
		\centering
		\includegraphics[width=\linewidth]{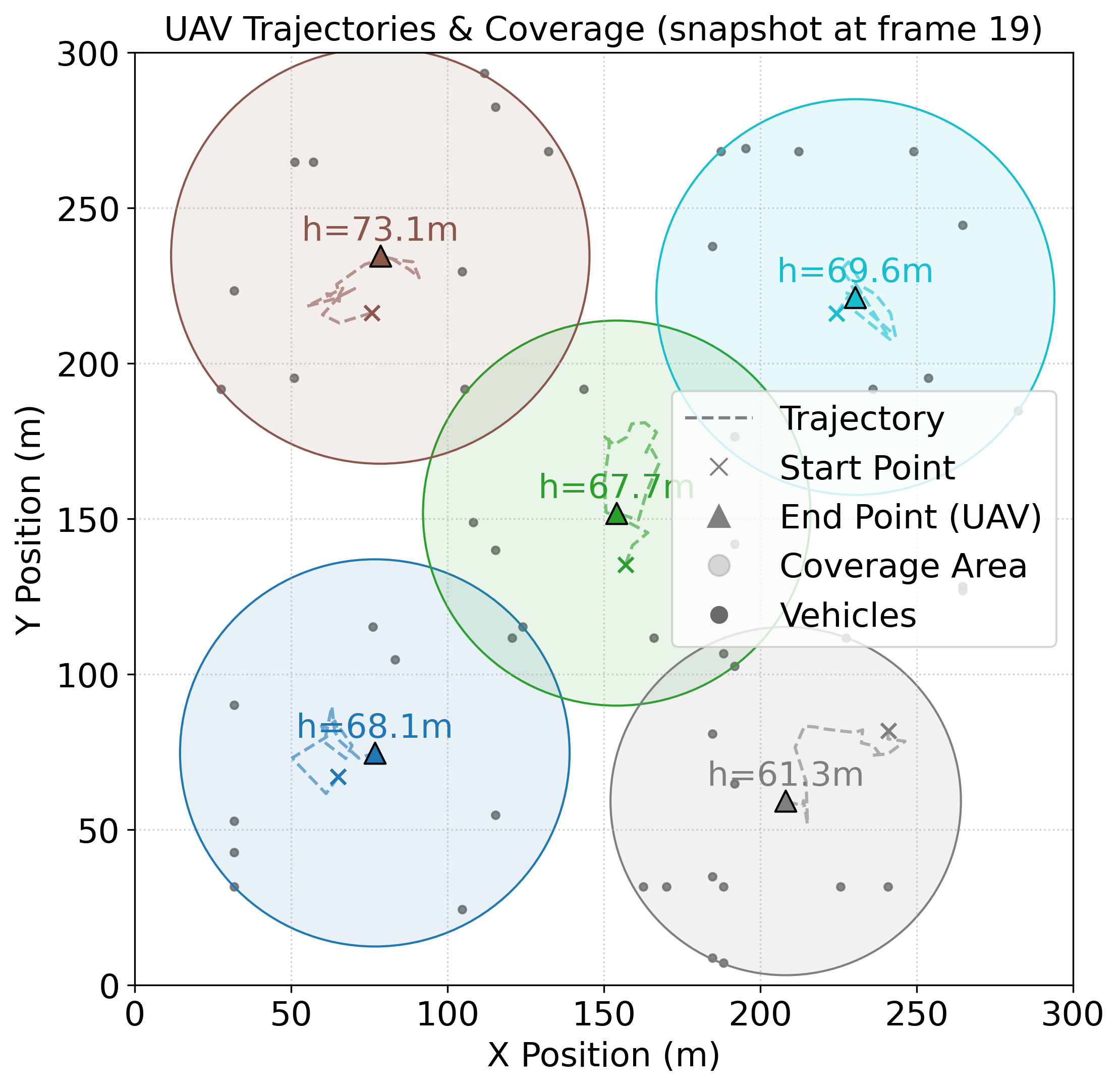}
		\caption{UAV historical movement trajectory}
		\label{fig:uav_trace}
	\end{minipage}
	\hfill
	\begin{minipage}{0.34\textwidth}
		\centering
		\includegraphics[trim=0cm 0cm 0cm 0cm, clip, width=\linewidth]{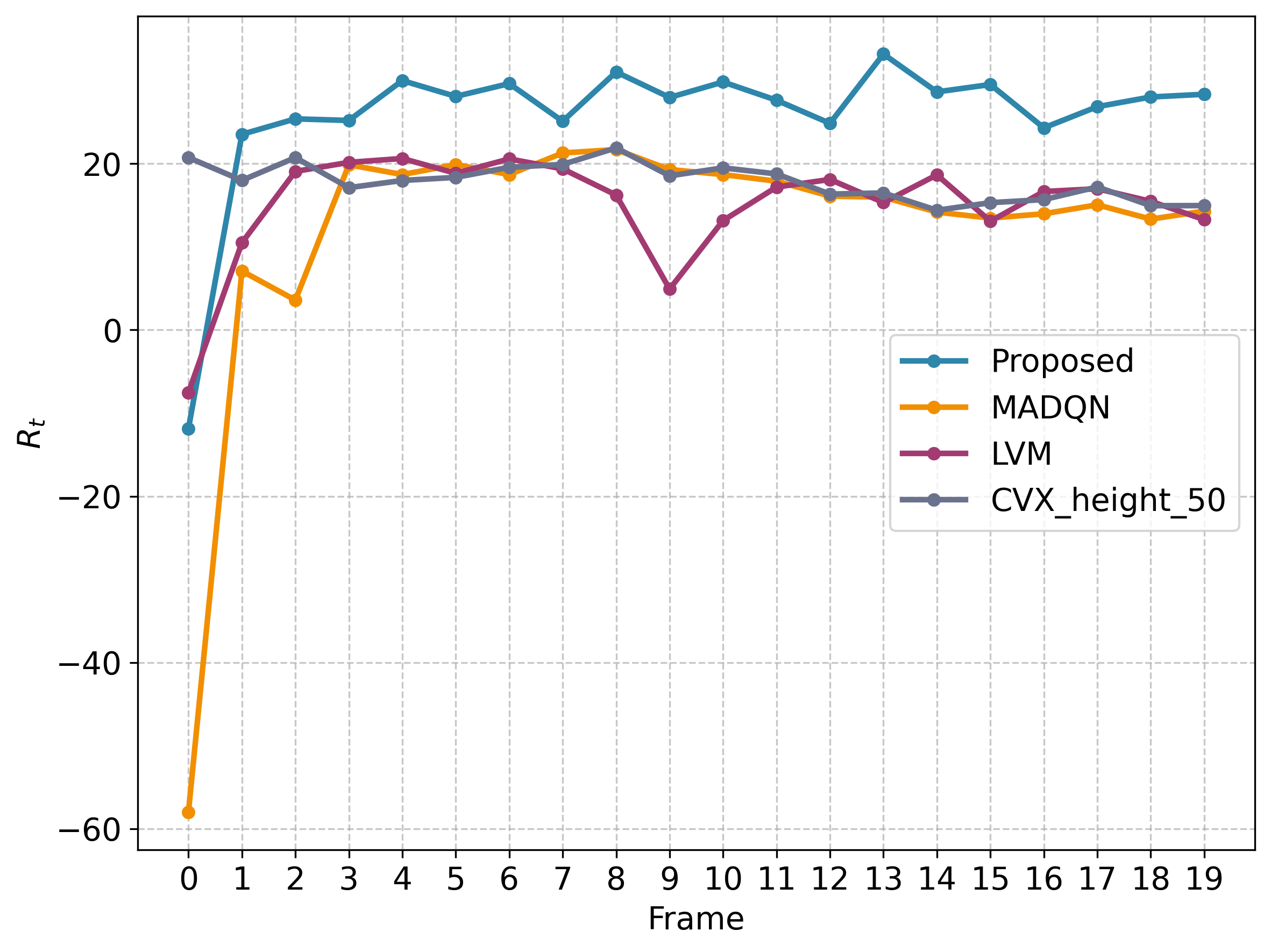}
		\caption{Average coverage metric comparison}
		\label{path_planning_performance}
	\end{minipage}
	\hfill
	\begin{minipage}{0.34\textwidth}
		\centering
		\includegraphics[trim=0cm 0cm 0cm 0cm, clip, width=\linewidth]{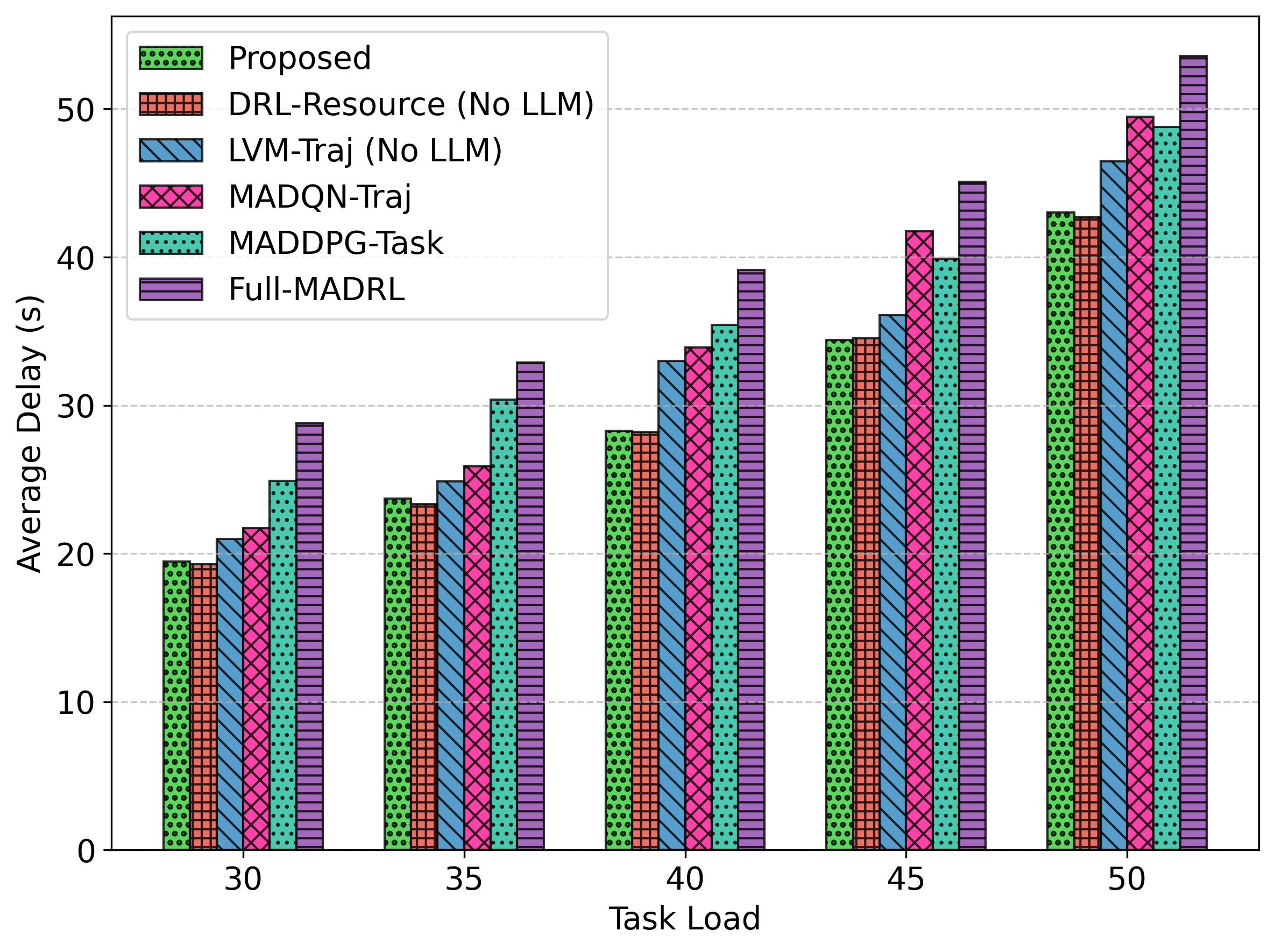}
		\caption{Delay vs. Load}
		\label{fig:task_completion_time}
	\end{minipage}
	
	\vspace{0.2cm}
\end{figure*}

\begin{figure*}[t]
	\centering
	\begin{minipage}{0.32\textwidth}
		\centering
		\includegraphics[trim=0cm 0cm 0cm 0.2cm, clip, width=\linewidth]{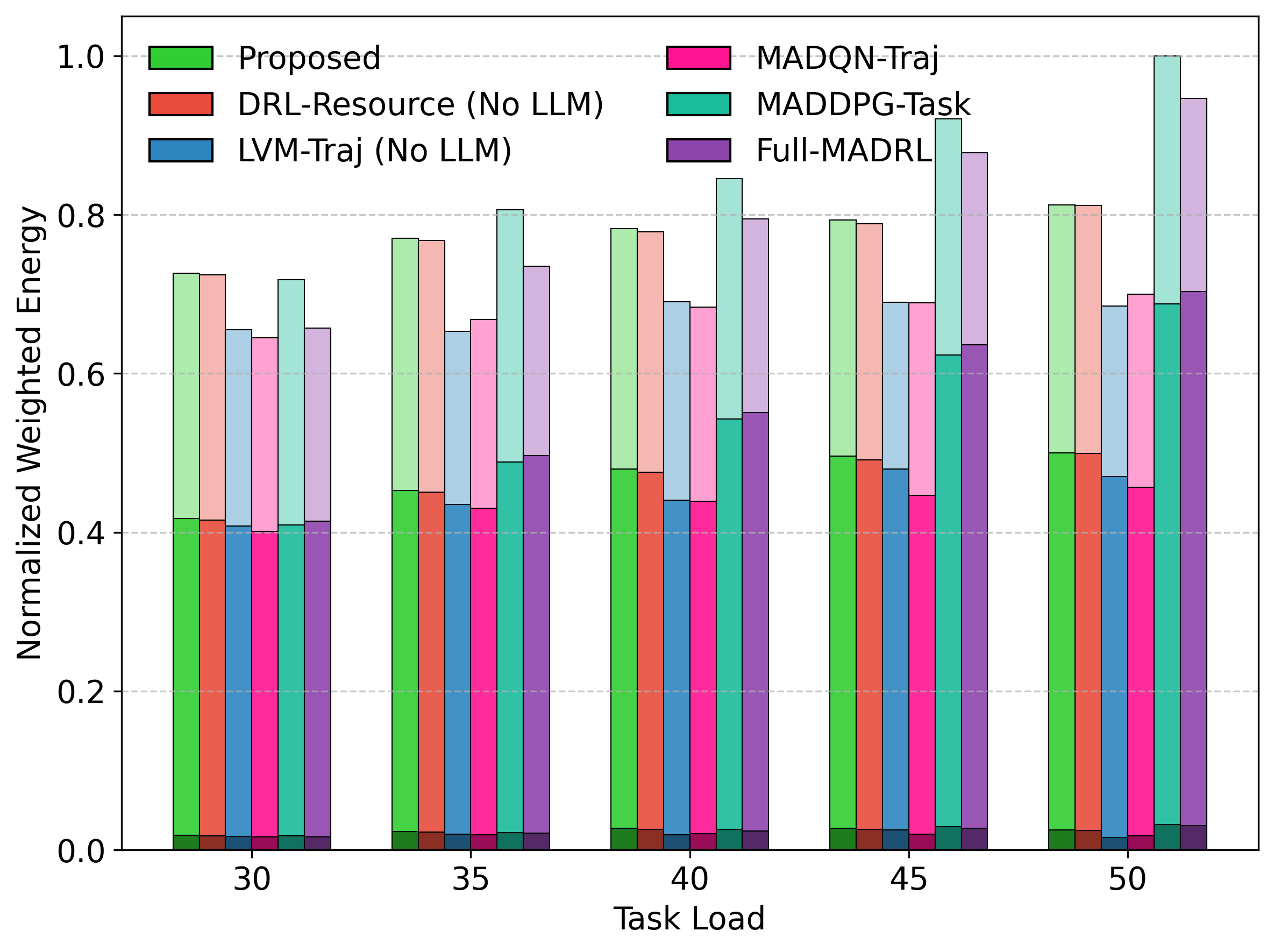}
		\caption{Energy vs. Load}
		\label{fig:energy_consumption}
	\end{minipage}
	\hfill
	\begin{minipage}{0.32\textwidth}
		\centering
		\includegraphics[width=\linewidth]{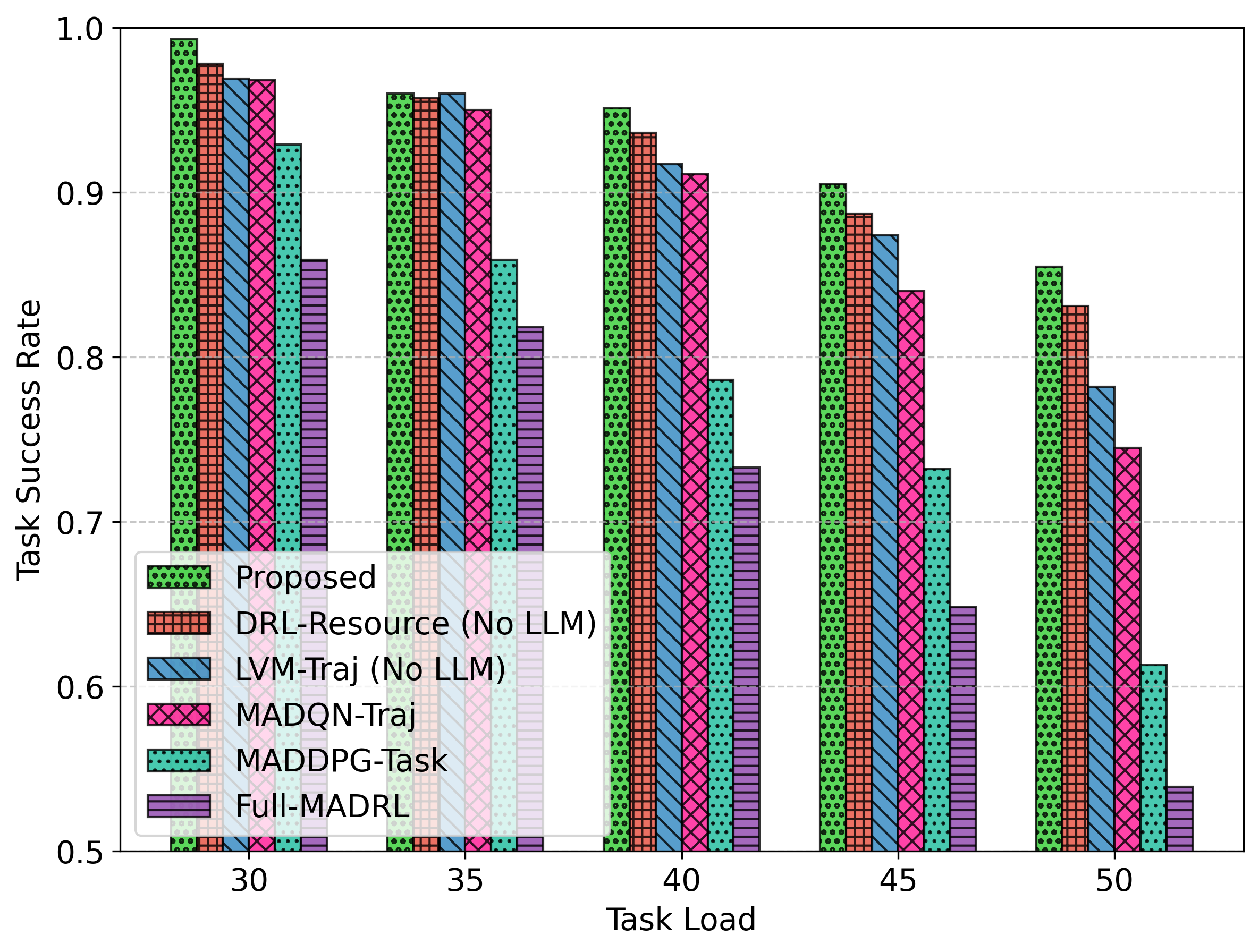}
		\caption{Success Rate vs. Load}
		\label{fig:task_success_rate}
	\end{minipage}
	\hfill
	\begin{minipage}{0.32\textwidth}
		\centering
		\includegraphics[width=\linewidth]{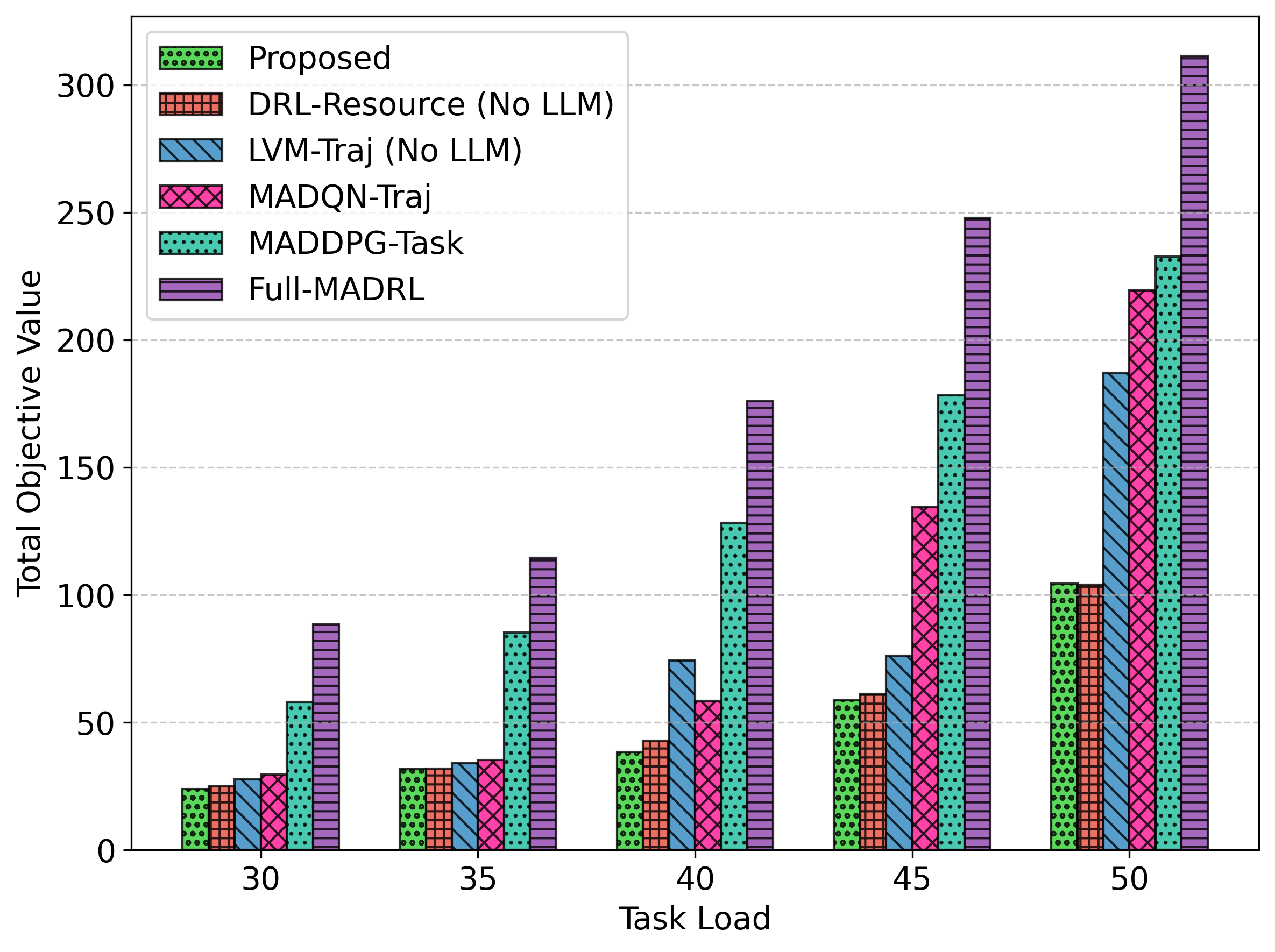}
		\caption{Total Objective vs. Load}
		\label{fig:object}
	\end{minipage}
\end{figure*}
\begin{figure*}[t]
	\centering
	\subfigure[vehicle allocation ratios]{
		\label{AoIOver_0inf}        \includegraphics[width=0.32\textwidth]{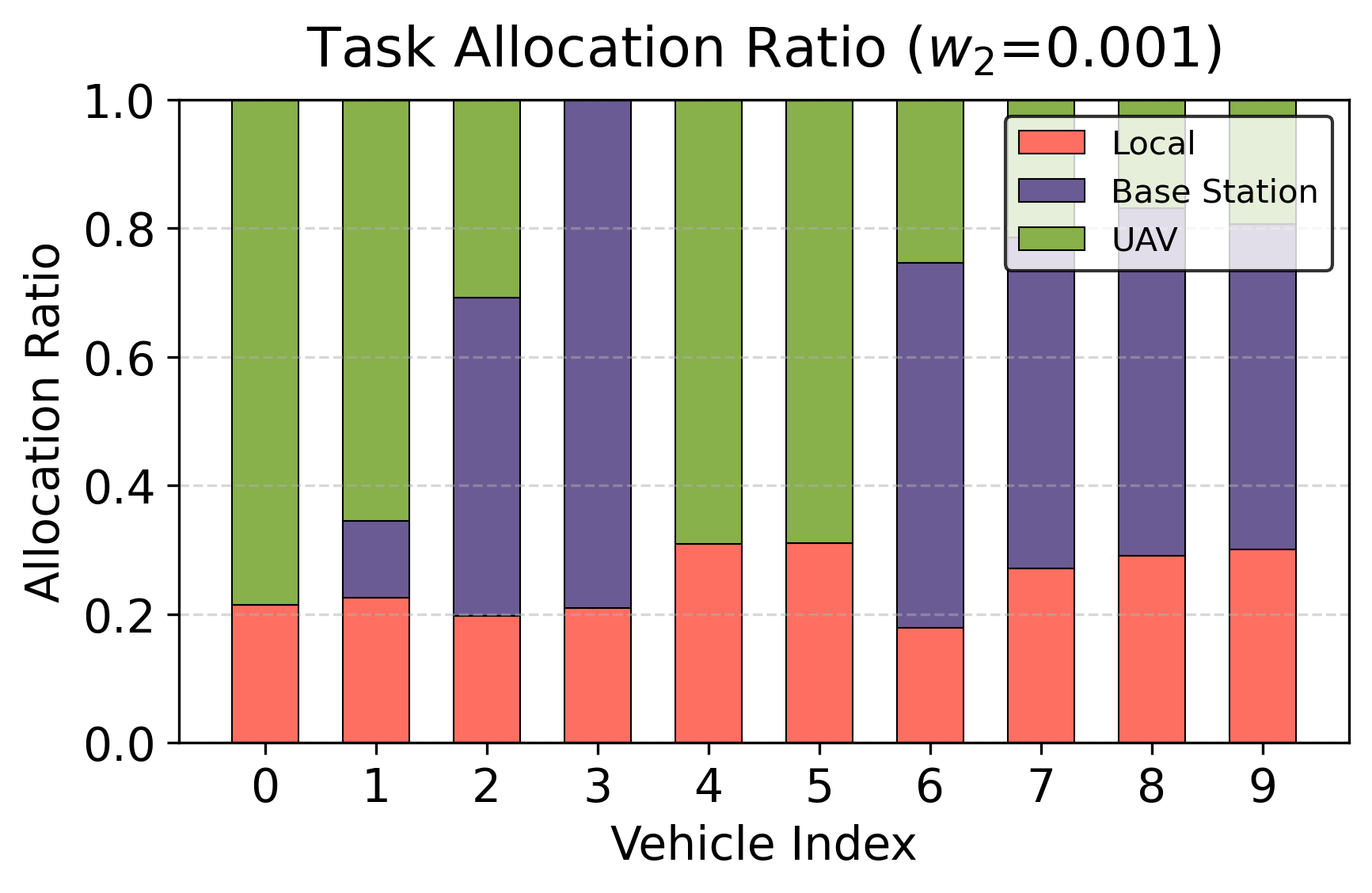}}
	\hspace{-0.4cm} 
	\subfigure[vehicle delay components]{
		\label{AoIOver_20inf}
		\includegraphics[width=0.32\textwidth]{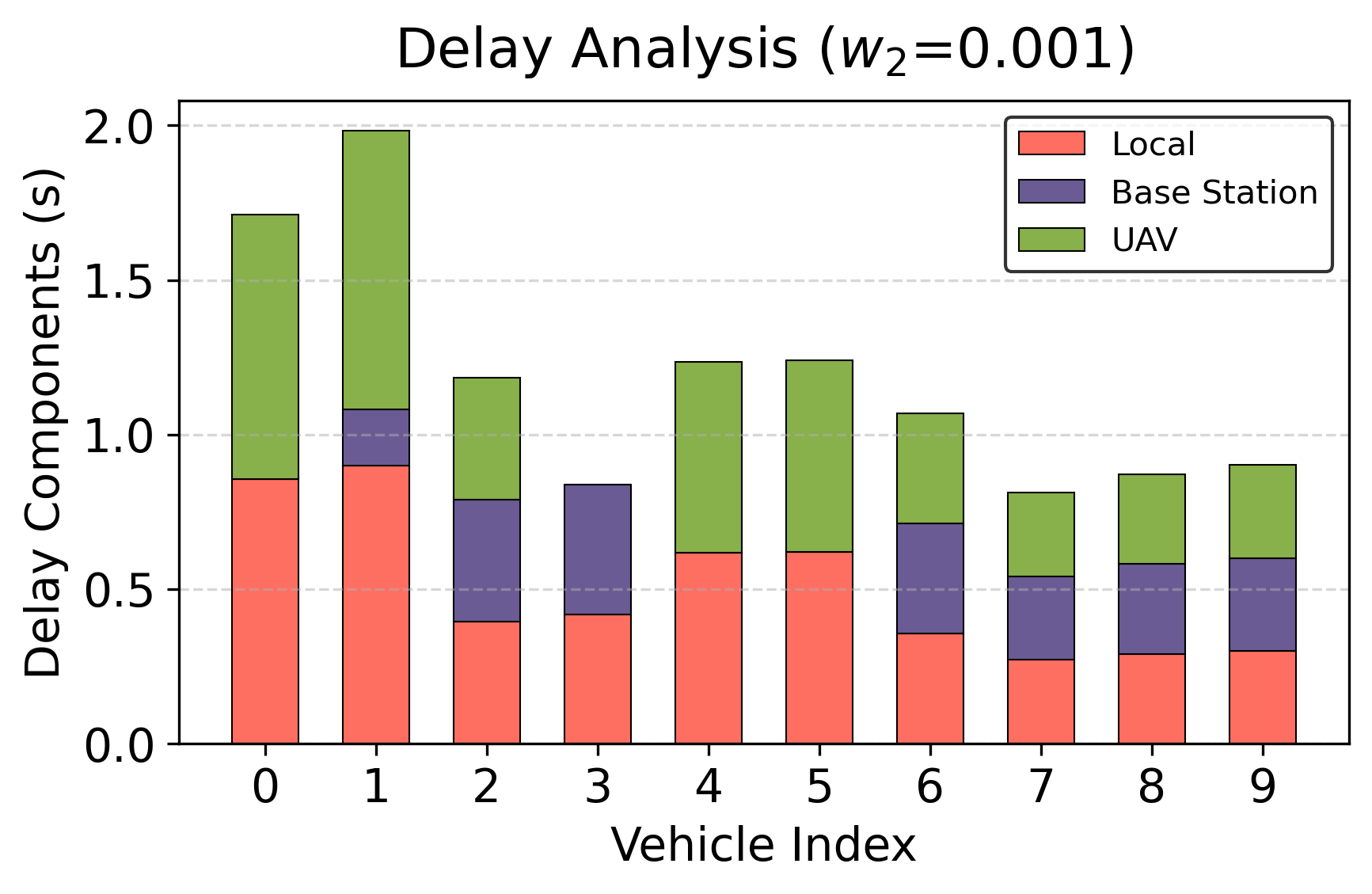}}
	\hspace{-0.4cm} 
	\subfigure[load distribution]{
		\label{AoIOver_40inf}
		\includegraphics[width=0.32\textwidth]{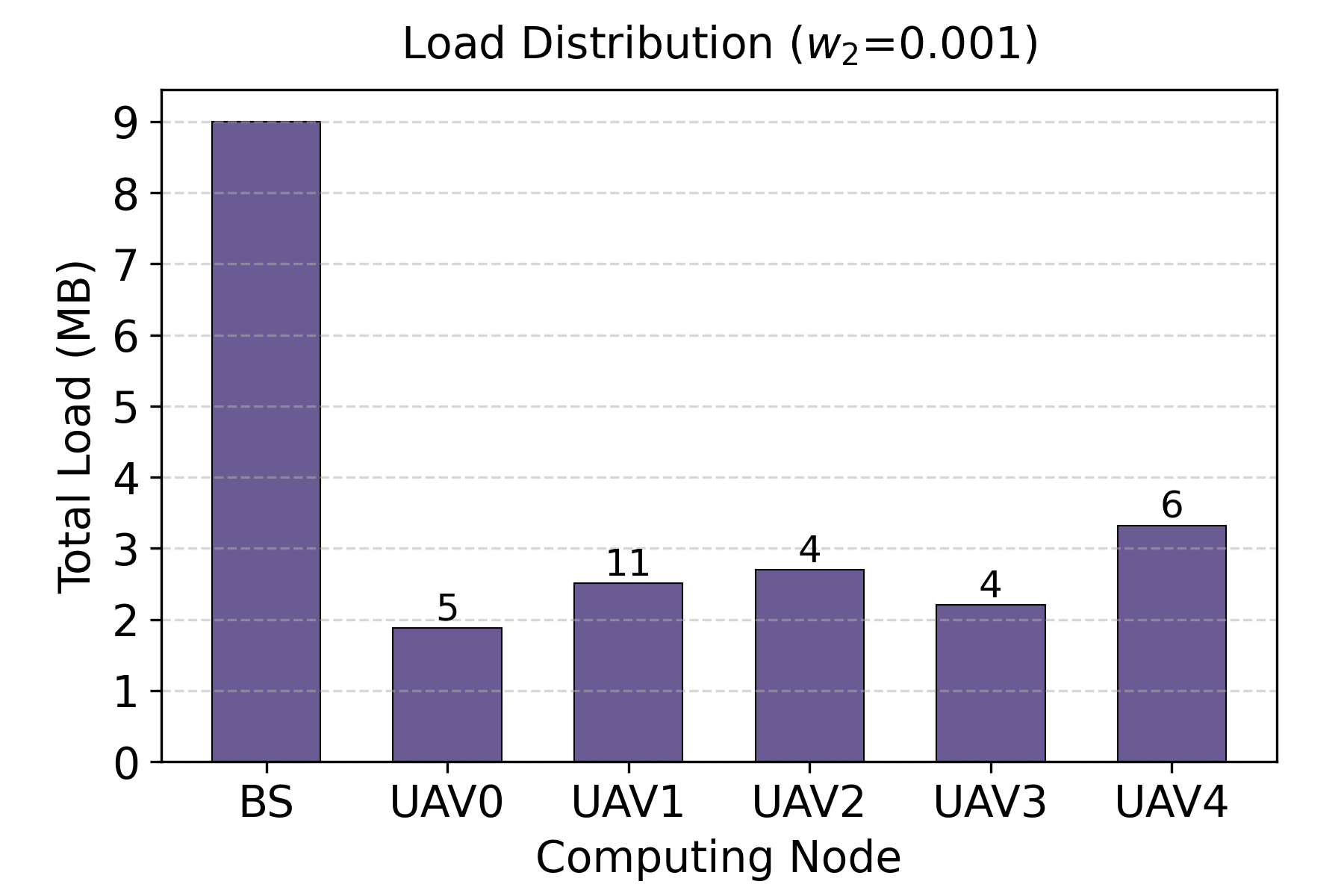}}
	\vspace{-0.2cm}
	\caption{Task Allocation Strategy with Emphasis on Delay}
	\label{AoIOver}
	\vspace{-0.2cm}
\end{figure*}
\begin{figure*}[t]
	\centering
	\subfigure[vehicle allocation ratios]{
		\label{DistOver_0inf}
		\includegraphics[width=0.32\textwidth]{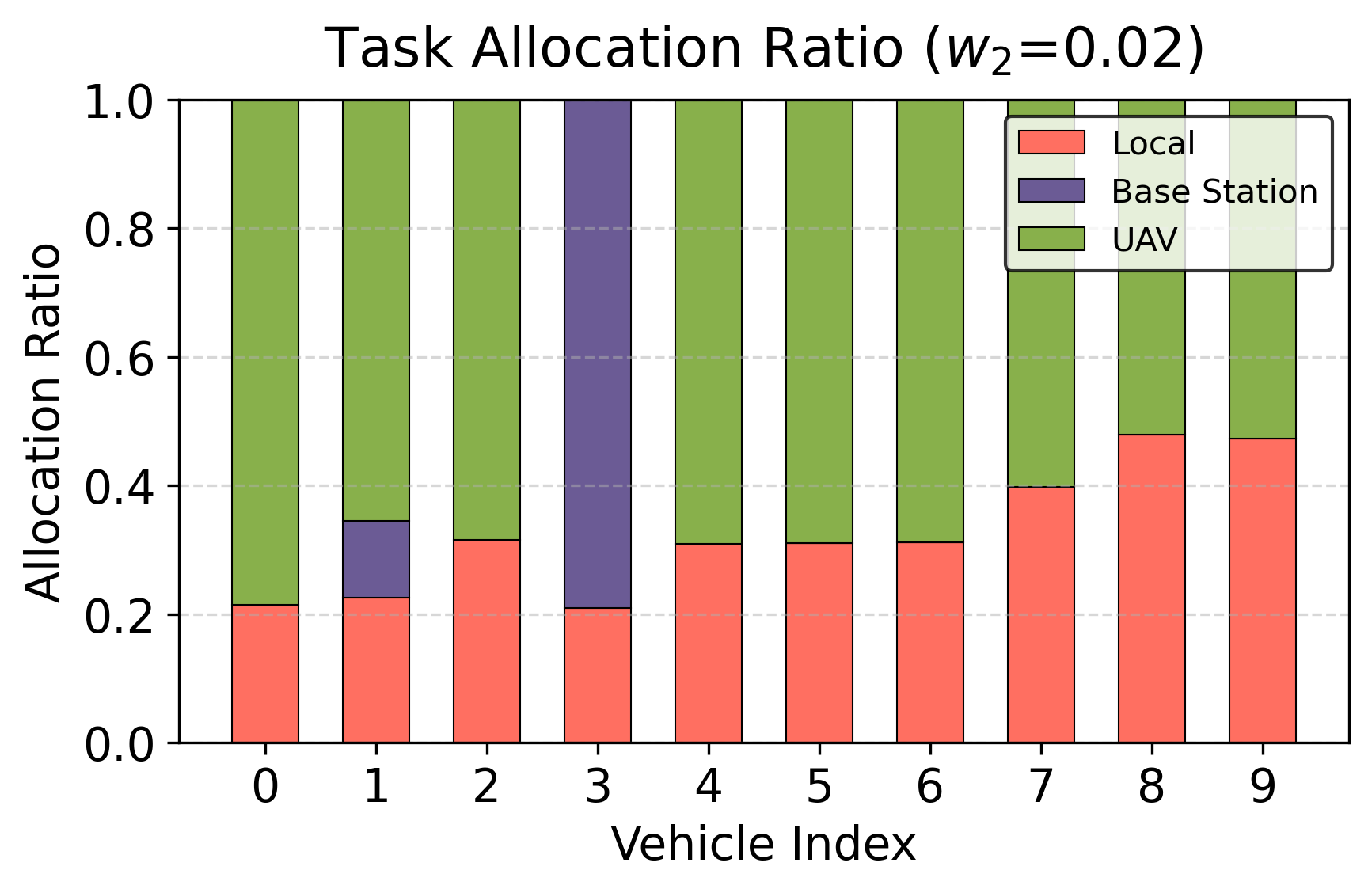}}
	\hspace{-0.4cm} 
	\subfigure[vehicle delay components]{
		\label{DistOver_20inf}
		\includegraphics[width=0.32\textwidth]{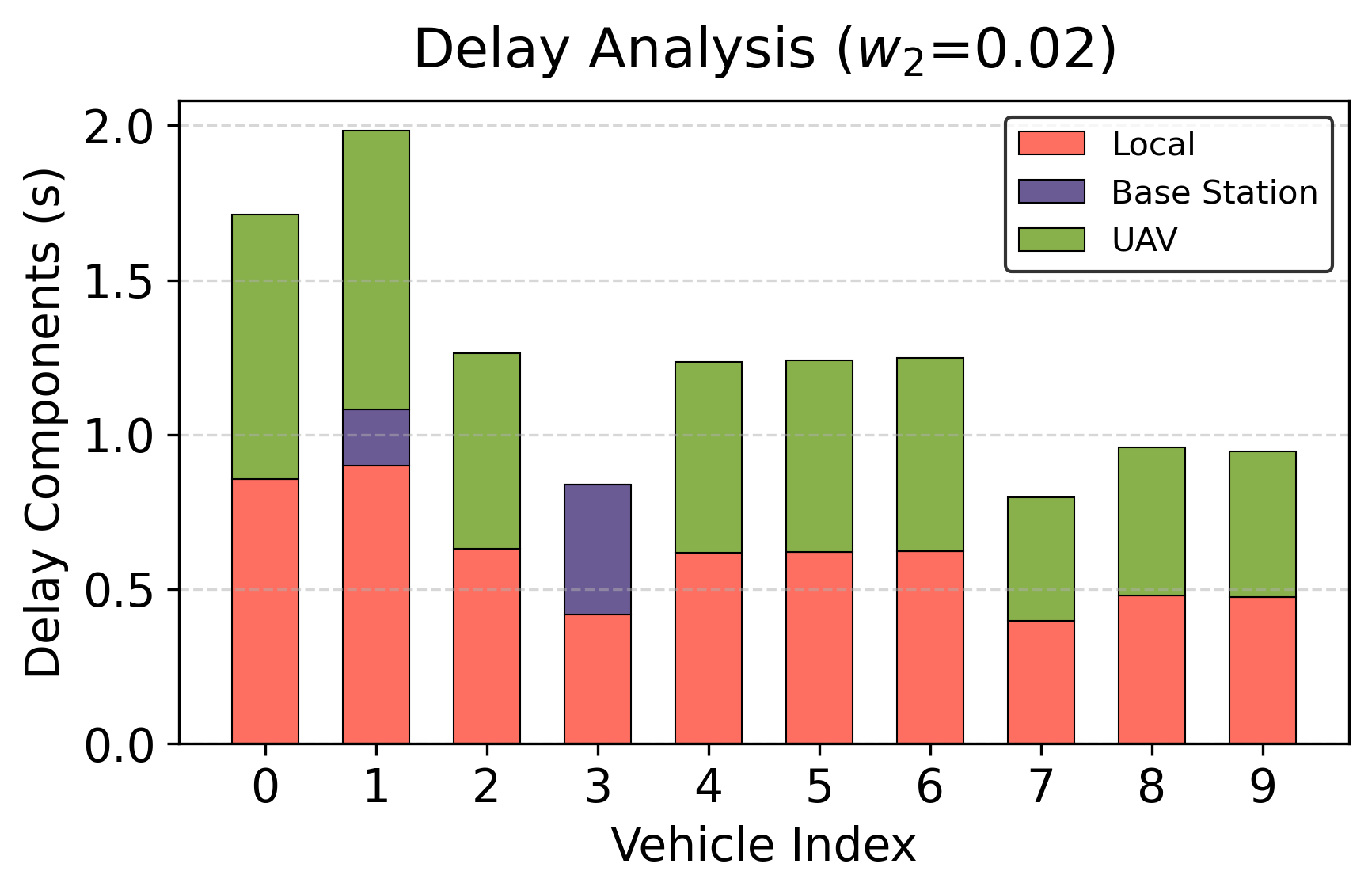}}
	\hspace{-0.4cm} 
	\subfigure[load distribution]{
		\label{DistOver_40inf}
		\includegraphics[width=0.31\textwidth]{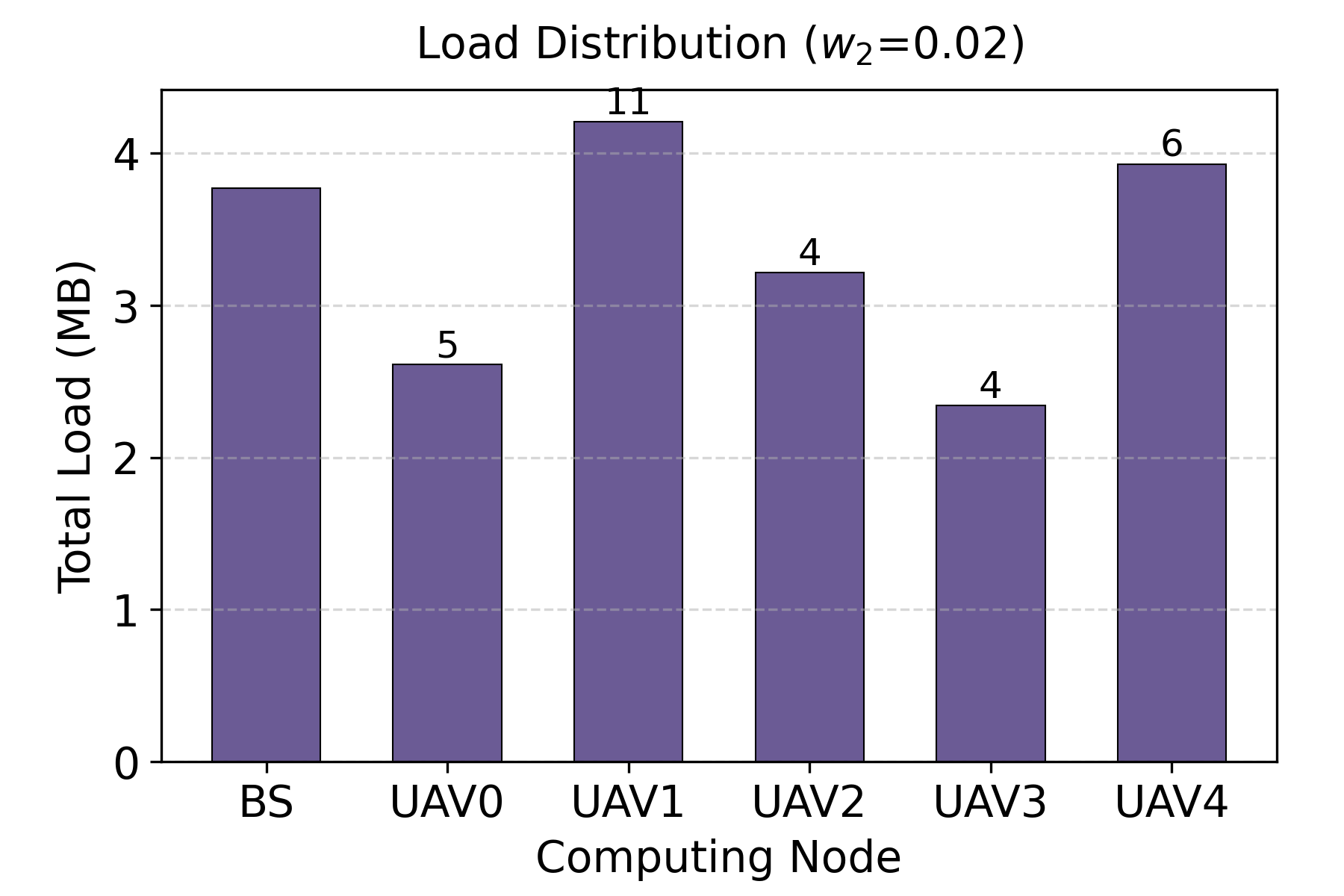}}
	\vspace{-0.2cm}
	\caption{Allocation Strategy with Emphasis on Energy Consumption}
	\label{DistOver}
	\vspace{-0.2cm}
\end{figure*}

Fig.~\ref{fig:task_completion_time} shows the average system task completion time under different total loads. The Proposed and DRL-Resource (No LLM) methods achieve the best performance. In the Proposed method, the LLM enhances fairness by reallocating communication resources for long-tail tasks, improving the task success rate. However, since this is a multi-objective problem, according to the Pareto principle, such adjustments inevitably lead to a slight increase in average delay compared to the DRL-Resource method. Trajectory planning methods based on LVM and MADQN perform slightly worse; as shown in Fig.~\ref{position_coverage}, their inferior coverage quality compared to CVX leads to higher communication delays. Among the modules, the task offloading scheme has the most significant impact on delay. The LP-based method precisely solves for optimal allocation, resulting in lower delay, whereas DDPG-based methods struggle to find the optimum due to learning bottlenecks, yielding higher task completion latencies.

Fig.~\ref{fig:energy_consumption} compares the weighted normalized energy consumption of different methods (Light color: flight; Medium: computation; Dark: communication; Weights: 0.3:1:100). MADQN and LVM methods exhibit lower flight energy consumption because they tend to make conservative altitude adjustments, sacrificing coverage precision for energy savings. The MADDPG-based task allocation method incurs higher computation energy consumption as it tends to offload more tasks to high-frequency base stations. In contrast, the LP-based method balances energy and delay more effectively.

Fig.~\ref{fig:task_success_rate} shows the average task success rate, defined as the proportion of tasks completed within the deadline. As the total load per slot increases, the success rate of all methods decreases due to the limited total computing capacity of the environment. When the load approaches the capacity limit, base stations and some UAVs reach saturation, forcing some vehicles to compute locally and resulting in timeouts. Notably, the Proposed method significantly improves the success rate by using the LLM to reallocate resources for failed and surplus tasks, allowing potential failures to be offloaded earlier. Furthermore, the success rates of MADDPG and Full-MADRL combinations drop sharply with increasing load. This is because MADDPG-based allocation fails to strictly enforce queue capacity constraints, leading to inevitable task failures upon overflow.

Fig.~\ref{fig:object} compares different methods in terms of the total objective function value~\eqref{eq:objective}. Given the multi-objective nature of the problem, LP-based methods perform better overall, as mathematical solvers are most effective at minimizing the Min-Max latency of components. The LVM-based method shows a sharp increase in objective value when the load reaches 40 Mb, corresponding to the rise in average task completion time in Fig.~\ref{fig:task_completion_time}. The Proposed method achieves a higher success rate, translating to a smaller penalty for delay violations, and thus a lower total objective value.
\vspace{-0.3cm}
\subsection{Analysis of Task Allocation Strategies}
\FigRef{AoIOver} and \FigRef{DistOver} illustrate the strategic differences in a single decision instance under varying weights for delay and energy consumption.
\FigRef{AoIOver} shows the allocation results when focusing solely on delay (energy weight $\approx 0$). As seen in Fig.~\ref{AoIOver}(c), the base station load quickly reaches its 9 Mb limit. Most vehicles prioritize allocating tasks to the base station; once full, tasks spill over to UAVs, and finally to local computation. In Fig.~\ref{AoIOver}(b), the completion times for different parts of most vehicle tasks are equal, aligning with the intuition of minimizing the maximum completion time (Make-span) by balancing loads across nodes.

\FigRef{DistOver} shows the results when energy consumption is heavily weighted. The proportion of tasks assigned to UAVs increases significantly (Fig.~\ref{DistOver}(c)), while the base station load decreases. Fig.~\ref{DistOver}(a) shows that most vehicles allocate tasks exclusively to UAVs, bypassing the base station. However, Vehicle 3, located outside the UAV coverage area, is forced to offload to the base station, demonstrating the algorithm's adaptability to topological constraints.

It is noteworthy that these experiments were conducted under a moderate load of 30 Mb. When the load increases to 50 Mb, the resulting strategies become identical regardless of weight settings. This is because, as the system approaches its capacity limit, the feasibility of balancing delay and energy is lost; all available computing resources must be fully utilized to avoid task failures.



\section{Conclusion} \label{conclusion}

\textcolor{red}{In this paper, we propose a joint optimization framework for 3D trajectory control, resource allocation, and task offloading in multi-UAV-assisted IoV systems. To address the coupling and non-convexity of the problem, we decompose it into three subproblems and solve them via a hierarchical execution flow. Specifically, a sequential distributed optimization algorithm based on SOCP is developed to optimize UAV trajectories under dynamic vehicle topologies. To overcome the generalization limitations of traditional DRL in long-tail scenarios, we introduce an LLM-based macro-scheduler within an alternating optimization loop. This framework synergizes the high-efficiency initial scheduling of DRL with the semantic reasoning capabilities of LLMs, enabling precise resource reallocation for failed and surplus tasks. Crucially, a reward decoupling mechanism is implemented to ensure the training stability of the DRL agent under external interventions. Simulation results demonstrate that the proposed method significantly outperforms baseline algorithms (e.g., MADRL, MADQN) in terms of task success rate, system latency, and energy efficiency. Furthermore, the integration of KV caching and MoE architecture ensures the feasibility of deploying large-scale models at the network edge. Future work will explore the coordination of heterogeneous UAV swarms and the integration of multi-modal LLMs for complex urban semantic environment perception.}

\ifCLASSOPTIONcaptionsoff
  \newpage
\fi



\end{document}